\documentclass[sigconf]{acmart}
\AtBeginDocument{%
  }

\copyrightyear{2026}
\acmYear{2026}
\setcopyright{cc}
\setcctype{by}
\acmConference[KDD '26]{Proceedings of the 32nd ACM SIGKDD Conference on Knowledge Discovery and Data Mining V.2}{August 09--13, 2026}{Jeju Island, Republic of Korea}
\acmBooktitle{Proceedings of the 32nd ACM SIGKDD Conference on Knowledge Discovery and Data Mining V.2 (KDD '26), August 09--13, 2026, Jeju Island, Republic of Korea}
\acmDOI{10.1145/3770855.3818075}
\acmISBN{979-8-4007-2259-2/2026/08}

\usepackage{multicol}
\usepackage{multirow}
\usepackage{xcolor}
\usepackage{colortbl}
\usepackage{enumitem}
\usepackage{hyperref}
\usepackage{fontawesome}
\usepackage[breakable]{tcolorbox}
\tcbuselibrary{listings,breakable}
\usepackage{CJKutf8}
\usepackage{quoting}
\usepackage{balance}

\newcommand{\TheName}{\texttt{APEX-SQL}}

\definecolor{my_red}{RGB}{192,0,0}
\definecolor{my_green}{RGB}{30, 150, 30}

\newtcolorbox[auto counter, number within=section]{promptbox}[1][]{
    colback=yellow!10,
    colframe=yellow!70!black,
    arc=10pt,
    title={\textcolor{black}{\textbf{\ifstrempty{#1}{Prompt \thetcbcounter}{#1}}}},
    colbacktitle=yellow!40,
    breakable,
}

\usepackage{enumitem}
\usepackage{listings}
\usepackage{csquotes}
\begin{document}

\title{\TheName{}: Talking to the Data via Agentic Exploration for Text-to-SQL}
\author{Bowen Cao}
\authornote{This work was done during an internship at Tencent LIGHTSPEED.}
\authornote{Both authors contributed equally to this research.}
\email{bwcao@link.cuhk.edu.hk}
\affiliation{%
  \institution{The Chinese University of Hong Kong}
  \city{Hong Kong}
  \country{China}
}

\author{Weibin Liao}
\authornotemark[2]
\email{liaoweibin@stu.pku.edu.cn}
\affiliation{%
  \institution{Peking University}
  \city{Beijing}
  \country{China}}

\author{Yushi Sun}
\authornote{Correspondence to: Yushi Sun and Dong Fang.}
% \authornote{Correspondence to: Yushi Sun (steveyssun@tencent.com) and Dong Fang (stanleyfang@tencent.com).}
\email{ysunbp@connect.ust.hk}
\affiliation{%
  \institution{LIGHTSPEED}
  \city{Shenzhen}
  \country{China}
}

\author{Dong Fang}
\email{df572@outlook.com}
\authornotemark[3]
\affiliation{%
  \institution{LIGHTSPEED}
  \city{Shenzhen}
  \country{China}
}

\author{Haitao Li}
\email{729156675@qq.com}
\affiliation{%
  \institution{LIGHTSPEED}
  \city{Shenzhen}
  \country{China}
}

\author{Wai Lam}
\email{wlam@se.cuhk.edu.hk}
\affiliation{%
  \institution{The Chinese University of Hong Kong}
  \city{Hong Kong}
  \country{China}}
  
\renewcommand{\shortauthors}{Bowen Cao et al.}

\begin{abstract}
Text-to-SQL systems powered by Large Language Models have excelled on academic benchmarks but struggle in complex enterprise environments. The primary limitation lies in their reliance on static schema representations, which fails to resolve semantic ambiguity and scale effectively to large, complex databases. To address this, we propose \TheName{}, an Agentic Text-to-SQL Framework that shifts the paradigm from passive translation to agentic exploration. Our framework employs a hypothesis-verification loop to ground model reasoning in real data. 
In the schema linking phase, we use logical planning to verbalize hypotheses, dual-pathway pruning to reduce the search space, and parallel data profiling to validate column roles against real data, followed by global synthesis to ensure topological connectivity. 
For SQL generation, we introduce a deterministic mechanism to retrieve exploration directives, allowing the agent to effectively explore data distributions, refine hypotheses, and generate semantically accurate SQLs.
Experiments on BIRD (70.65\% execution accuracy) and Spider 2.0-Snow (53.03\% execution accuracy) demonstrate that \TheName{} outperforms competitive baselines. 
Further analysis reveals that agentic exploration acts as a performance multiplier, unlocking the latent reasoning potential of foundation models in enterprise settings. Ablation studies confirm the critical contributions of each component in ensuring robust and accurate data analysis.
Our code is released at \url{https://github.com/Tencent/APEX-SQL-Project}.
\end{abstract}





\maketitle

% —— 加在第一页栏底的资助声明 ——
\renewcommand{\thefootnote}{}\footnotetext{%
  This research/paper was partially supported by the Center for Perceptual and Interactive Intelligence (CPII) Ltd. under the Innovation and Technology Commission’s InnoHK scheme.%
}
\renewcommand{\thefootnote}{\arabic{footnote}}  % 还原编号，避免影响后续脚注

\begin{figure}[t]
    \centering
    \includegraphics[width=\linewidth]{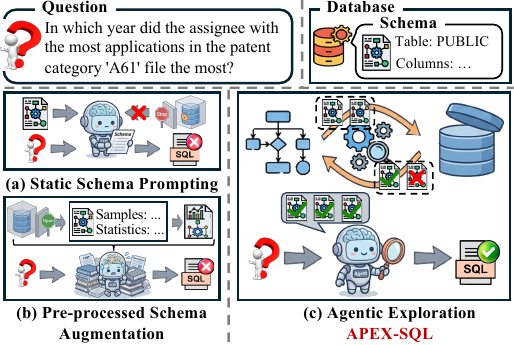}
    % \vspace{-5mm}
    \caption{A comparison of Text-to-SQL paradigms. (a) \textbf{Static Schema Prompting} relies solely on the schema, causing hallucinations when metadata is ambiguous. (b) \textbf{Pre-processed Schema Augmentation} attempts to enrich context with preliminary data profiling, yet introduces irrelevant noise and token overhead. (c) \TheName{} introduces \textbf{Agentic Exploration}, employing a hypothesis-verification loop to actively interrogate the database and ground logical reasoning.}
    % \vspace{-5mm}
    \label{fig:intro}
\end{figure}
\section{Introduction}

Text-to-SQL~\cite{shi2025survey,liu2025survey} aims to make data analysis accessible by translating natural language questions into executable SQL queries. Recent advancements in Large Language Models (LLMs) have significantly improved execution accuracy on academic benchmarks like Spider 1.0~\cite{yu2018spider} and BIRD~\cite{li2023can}. However, this performance drops substantially in real-world enterprise environments~\cite{lei2024spider}.

A fundamental limitation of current approaches lies in their reliance on passive perception.
Standard methods depend on static schema definitions (Figure~\ref{fig:intro}a), but enterprise databases often feature opaque column names, hiding true semantics within data values rather than metadata. Without access to the data, models are forced to guess.
To address this, existing works attempt to inject pre-processed data profiles (Figure~\ref{fig:intro}b). However, this approach has limitations. On one hand, static summaries remain detached from specific question contexts, failing to provide the exact evidence needed. On the other hand, covering the vast scale of databases introduces noise and incurs heavy token overhead.
Consequently, these passive paradigms disconnect reasoning from reality, leading to queries that are syntactically correct but semantically flawed.

To resolve this, we propose a paradigm shift from passive perception to agentic exploration (Figure~\ref{fig:intro}c). Central to our approach is a Hypothesis-Verification loop where the agent formulates logical assumptions based on the user question and validates them by executing exploratory SQLs against actual data. This process transforms Text-to-SQL from a static translation task into an interactive reasoning task grounded in database reality.

We introduce \TheName{}, a unified framework applying agentic exploration to both Schema Linking and SQL Generation. 
First, given massive industrial databases, Schema Linking is crucial to isolate the relevant schema elements before processing. 
We employ Logical Planning to verbalize question requirements into schema-agnostic hypotheses, and introduce Dual-Pathway Pruning to compress the search space, making the subsequent validation feasible within large databases. The agent then conducts Parallel Data profiling to validate these hypotheses against real data, followed by Global Synthesis to ensure topological connectivity across the verified elements. 
For SQL Generation, we bridge the gap between abstract intent and data-level constraints through Deterministic Guidance Retrieval, which maps logical steps into exploration directives.
Equipped with this guidance, the agent performs agentic exploration by navigating a flexible action space to profile data distributions, consolidate findings, and synthesize candidate queries. The process concludes with a final confirmation step to verify semantic fidelity against the accumulated evidence, ensuring the generated SQL is syntactically and semantically correct.

Extensive experiments on BIRD and Spider 2.0 validate this paradigm shift. On \textbf{BIRD-Dev}, \TheName{} achieves an execution accuracy of 70.7\%, surpassing competitive baselines such as OpenSearch-SQL (69.3\%)~\cite{xie2025opensearchsqlenhancingtexttosqldynamic} and RSL-SQL (67.2\%)~\cite{cao2024rsl}.
On the enterprise-grade \textbf{Spider 2.0-Snow}, \TheName{} attains 53.0\% execution accuracy, outperforming existing agentic systems~\cite{hao2025text,deng2025reforce}. Further analysis reveals that agentic exploration acts as a performance multiplier, yielding up to an 18.33\% absolute gain for high-capacity models like DeepSeek-V3.2~\cite{liu2024deepseek}. 
Test-time compute scaling evaluations show that our framework effectively unlocks the latent reasoning potential of foundation models.
Moreover, ablation studies demonstrate that key components such as logical planning and exploration are crucial for improving schema linking performance; deterministic guidance in SQL generation enhances exploration effectiveness and efficiency, allowing the agent to resolve implementation uncertainties in fewer interaction rounds.

Our contributions can be summarized as follows:
\begin{itemize}
[leftmargin=*,itemsep=0pt,parsep=0.3em,topsep=0.3em,partopsep=0.3em]
    \item \textbf{Agentic Exploration for Schema Linking.} We propose a schema linking approach utilizing logical planning to verbalize hypotheses, dual-pathway pruning to reduce the massive search space, and parallel data profiling followed by global synthesis to ground column selection in empirical evidence.
    \item \textbf{Agentic Exploration for SQL Generation.} We introduce an agentic SQL generation process where deterministic retrieval provides exploration directives for the agent to verify hypotheses and consolidate data findings, autonomously resolving implementation ambiguities before query synthesis.
    \item \textbf{Empirical Validation \& Insights.} We demonstrate superior performance on BIRD and Spider 2.0 benchmarks, outperforming competitive baselines. Our analysis confirms that agentic exploration serves as a performance multiplier and is a highly effective mechanism for robust data analysis in enterprise environments.
\end{itemize}
% % \vspace{-0.5mm}
\section{Related Work}
% \paragraph{LLM-based Text-to-SQL with Schema Prompting.}
\paragraph{Static Schema Prompting.}
To empower LLMs for Text-to-SQL tasks, standard approaches typically linearize database schemas and supply them alongside queries as prompts~\cite{dong2023c3, pourreza2023din, liao2025learnat}. While methods such as DIN-SQL~\cite{pourreza2023din} and DAIL-SQL~\cite{gao2023text} optimize prompt structures to enhance reasoning capabilities, a fundamental challenge remains: LLMs are inherently ``blindness'' to underlying data distribution. Real-world databases often contain ambiguous column names, opaque abbreviations, or domain-specific codes that are semantically indecipherable without access to actual data~\cite{li2023can, sun2023reca, sun2026lakehopper}. Consequently, standard prompting confines models to surface-level schema representations, often inducing hallucinations when correct SQL logic depends on concrete data content rather than metadata alone. 
To mitigate the uncertainty arising from this ``blindness'', several works~\cite{lee2025mcs,pourreza2024chase,talaei2024chess} generate multiple candidate SQL queries and employ consensus or ranking-based strategies to select the most plausible output. 
\textit{However, these passive perception approaches still compel models to infer implementation details from ambiguous schemas. This severance of user intent from data reality frequently results in syntactically valid yet semantically flawed queries, particularly when the correct logic hinges on concealed data distributions.}

% \vspace*{-1.5mm}
\paragraph{Pre-processed Schema Augmentation.}
To bridge the gap between abstract schema definitions and concrete data values, recent research has introduced pre-processing pipelines to enrich schema representations. Approaches such as AskData~\cite{shkapenyuk2025automatic} leverage data profiling and query log analysis to statically augment schemas with value distributions and historical join paths prior to inference. Similarly, TA-SQL~\cite{qu2024before} employs LLMs to generate detailed natural-language descriptions for each database field. More broadly, knowledge-enhanced retrieval methods~\cite{sun2025kerag, sun2026cacherag} have explored augmenting language models with structured external knowledge to improve question answering accuracy. While these methods provide richer context, they rely on ``pre-computed'' knowledge that is static and often misaligned with the specific reasoning trajectory of a novel query. 
\textit{Despite offering additional context, these static augmentation strategies incur substantial token overhead and introduce irrelevant noise. By decoupling context retrieval from the evolving reasoning demands of complex queries, they fail to provide the targeted, dynamic verification necessary for accurate problem-solving.}

\begin{figure*}[t]
    \centering
    \includegraphics[width=\linewidth]{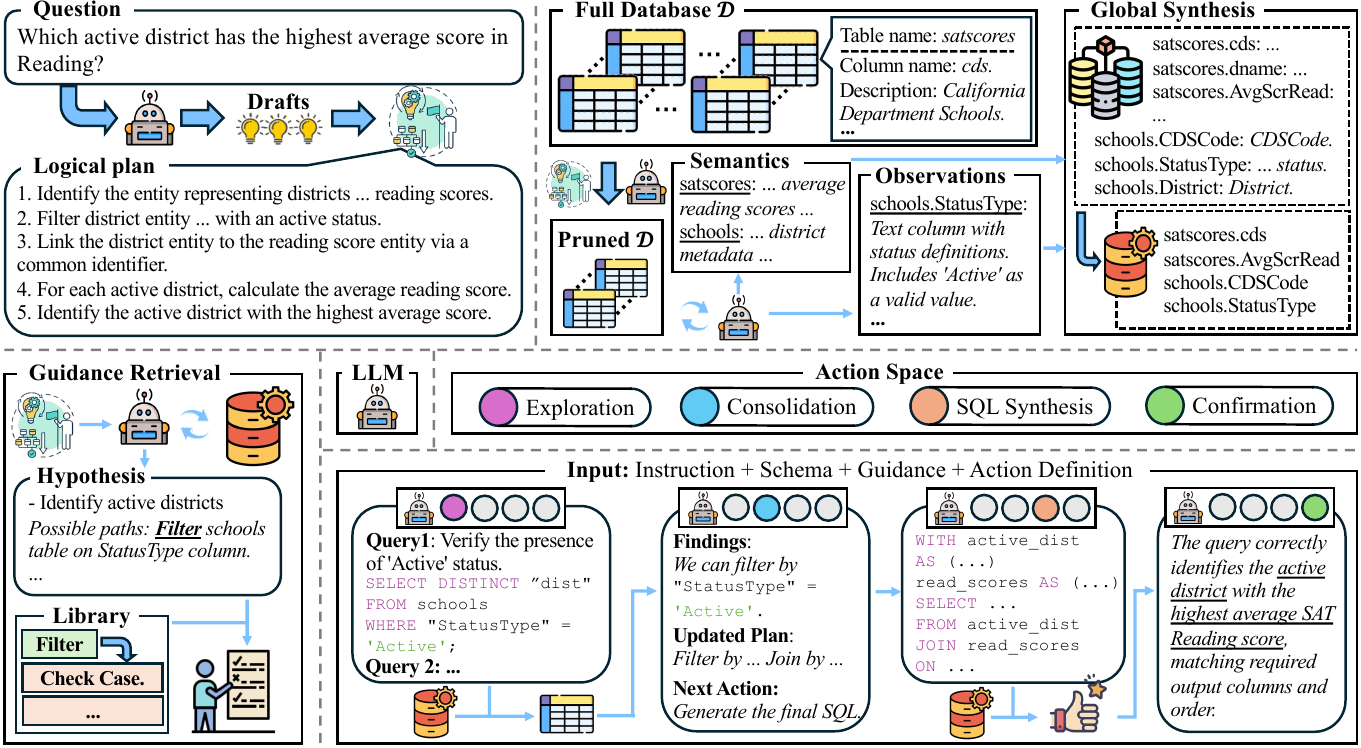}
    \caption{The proposed \TheName{} framework. \textbf{Schema Linking (Top):} To navigate massive databases, \TheName{} first verbalizes a logical plan, serving as a grounded reference for subsequent pruning and verification. It actively verifies the semantic roles of candidate tables through parallel profiling and summarizes key observations for each column. Finally, \TheName{} refines the schema subgraph by ensuring topological connectivity and recovering missing dependencies via global synthesis. \textbf{SQL Generation (Bottom):} \TheName{} employs a deterministic retrieval mechanism to map logical operations to specific exploration guidance. Directed by these constraints, it navigates a flexible action space to autonomously profile data distributions, consolidate exploration findings, synthesize candidate queries, or perform a final confirmation check to validate the executable SQL.}
    \label{fig:framwork}
% \vspace*{-3.5mm}
\end{figure*}

% \vspace*{-1.5mm}
\paragraph{Agentic Exploration.}
Contemporaneous literature have begun to adopt agentic workflows for database interaction. For instance, DSR-SQL~\cite{hao2025text} utilizes a dual-state mechanism to manage context updates, while AutoLink~\cite{wang2025autolink} focuses on iterative schema exploration to resolve linking issues in massive databases. Furthermore, ReFoRCE~\cite{deng2025reforce} introduces a column exploration mechanism to verify data values before generation. Although these methods mark a paradigm shift towards active problem-solving, they often treat schema linking and value verification as isolated sub-tasks or rely on unguided exploration, leading to computational inefficiency. 
\textit{Crucially, existing agentic frameworks lack a unified mechanism to discipline the exploration process. Consequently, they often suffer from inefficient, random probing trajectories that fail to systematically bridge the gap between abstract logical planning and concrete data-level constraints through a coherent hypothesis-verification loop.}
% \vspace*{-3.5mm}
\section{Methodology}
% This section outlines our approach to \textit{schema linking} and \textit{SQL generation}, presenting a novel agentic exploration framework that enables LLMs to dynamically interact with large-scale databases.

This section details how \TheName{} applies agentic exploration to schema linking and SQL generation through a hypothesis-verification loop. Figure \ref{fig:framwork} illustrates the framework of \TheName{}.

% \vspace*{-1.5mm}
\subsection{Problem Formulation}

Let $q$ denote a natural language question. The target database is structured as a set of tables $\mathcal{D} = \{T_1, \dots, T_m\}$\footnote{To simplify notation, $\mathcal{D}$ includes both schema and data; real data is accessed only during agentic exploration.}, where each table $T_j = \{c_1, \dots, c_k\}$ consists of columns along with their respective metadata (e.g., types and descriptions). \TheName{} addresses two sequential objectives. The first is \textit{schema linking}, which aims to identify a sufficient column set $\mathcal{D}^* \subset \mathcal{D}$ that maximizes the recall of columns relevant to the question while minimizing the number of columns. The second is \textit{SQL generation}, which focuses on synthesizing a correct query requested by $q$ against $D^*$.

% % \vspace*{0.5mm}
\noindent\textbf{\underline{H}ypothesis-\underline{V}erification (H-V).} 
% Ambiguous naming in enterprise databases obscures true semantics, necessitating a direct, data‑driven interpretation. To achieve agentic exploration within large-scale schemas, we propose a \textit{hypothesis–verification} loop as the foundation for both schema linking and SQL generation, where the system actively interacts with the database to refine its understanding. We will detail the specific steps of this process in the following sections.
Ambiguous naming in enterprise databases often obscures true semantics, necessitating a direct and data-driven interpretation.
To enable effective agentic exploration over large-scale schemas, we introduce a unified \textit{hypothesis–verification} loop, in which the system actively interacts with the database to iteratively refine its understanding.
Importantly, this agentic exploration is explicitly instantiated at two critical stages of the Text-to-SQL pipeline: \emph{schema linking} and \emph{SQL generation}.
We detail the design of agentic exploration for schema linking and SQL generation separately in the following subsections.

% \vspace*{-1.5mm}
\subsection{Agentic Exploration for Schema Linking}

% Accurately identifying the relevant schema subgraph is the cornerstone of text-to-SQL systems.
% Traditional approaches predominantly rely on \textit{static retrieval}, where models match query terms to table and column names based on lexical and semantic similarity. While effective for simple schemas, this paradigm struggles in large-scale enterprise databases characterized by ambiguous naming conventions and complex topological structures.

% To address this, we propose an \textbf{Active Schema Linking} framework. We reframe schema linking as an iterative \textit{hypothesis-verification} process where the system dynamically interrogates the database to ground user intent in empirical evidence. This approach progressively narrows the search space through three stages involving logical planning, dual-pathway pruning, and agentic verification.

% To navigate large-scale enterprise databases with ambiguous naming conventions and complex structures, we first apply $T^3D$ (\underline{T}alking \underline{T}o \underline{T}he \underline{D}ata) through Schema Linking via Agentic Exploration. We begin by generating logical plans to formulate schema-agnostic hypotheses, which are later validated through Evidence-Based Verification, grounded in the actual schema. To streamline the verification process, we introduce Dual-Pathway Pruning beforehand. The following sections provide a detailed breakdown of these steps.

To navigate large-scale enterprise databases with ambiguous naming conventions and complex structures, \TheName{} instantiates \emph{agentic exploration within the schema linking stage} through the H-V loop.
% $T^3D$ (\underline{T}alking \underline{T}o \underline{T}he \underline{D}ata).
Specifically, \TheName{} generates schema-agnostic hypotheses in the form of logical plans, which are subsequently grounded and validated against the database.
To further streamline this verification process, \TheName{} introduces a Dual-Pathway Pruning mechanism prior to verification.
The following sections provide a detailed breakdown of each component (Figure \ref{fig:framwork} (Top)).

% \vspace*{-1.5mm}
\subsubsection{Hypothesis Generation via Logical Planning}
\label{sec: logical plan}
% Directly mapping a query to a massive schema forces the model to perform retrieval and reasoning simultaneously, often leading to suboptimal outcomes due to the oversight of implicit intermediate steps and \textit{schema bias}, where LLMs hallucinate connections based on superficial string matching. To mitigate this, \TheName{} introduces a \textbf{Logical Planning} stage. Specifically, \TheName{} generates a schema-agnostic plan that hypothesizes the necessary computational steps and hidden constraints before the model is exposed to the noise of physical table names. 
Directly mapping a query onto a massive schema forces the LLM to conflate retrieval and reasoning, often resulting in suboptimal outcomes. This coupling obscures implicit intermediate steps and exacerbates \textit{schema bias}, wherein LLMs hallucinate spurious connections based on superficial string-level similarities.
To address this issue, \TheName{} introduces a dedicated \textbf{Logical Planning} stage.
Specifically, \TheName{} first generates a schema-agnostic plan that hypothesizes the required computational steps and latent constraints, deliberately isolating high-level reasoning from the noise introduced by concrete table and column names.

To ensure the derived plan covers all potential logical branches required to answer the query, \TheName{} employs a consensus-based synthesis strategy where multiple reasoning paths are aggregated into a unified master plan. Formally, \TheName{} generates $N$ independent candidates $\{P^{(i)}\}_{i=1}^N$ where each candidate is sampled from $p_\theta(\cdot \mid q, F^{sl}_{plan})$. These candidates are then integrated into a master plan $P^*$ through a secondary inference process defined by

% \vspace{-3mm}
\begin{equation}\label{equ:plan_agg}
    P^* = p_\theta \left( \cdot \mid q, \{P^{(1)}, \dots, P^{(N)}\}, F^{sl}_{agg} \right),
% \vspace{-1mm}
\end{equation}
% where $p_\theta$ represents the large language model and the instruction templates $F^{sl}_{plan}$ and $F^{sl}_{agg}$ are provided in \textcolor{red}{Appendix \ref{}}.
with $p_\theta$ the LLM and $F^{sl}_{\text{plan}}$, $F^{sl}_{\text{agg}}$ instruction templates.

% \vspace{-1.5mm}
\subsubsection{Search Space Reduction via Dual-Pathway Pruning}
\label{sec:pruning}
% Exhaustive data exploration is computationally prohibitive, necessitating a preliminary filtration step. However, standard relevance filtering is prone to false negatives, often discarding foreign keys that lack both lexical and semantic alignment with the query terms. To address this, \TheName{} employs a \textbf{Dual-Pathway Pruning} strategy. \TheName{} simultaneously execute a negative pass to remove obvious noise and a positive pass to identify relevant items, with the results later fused together.
% Formally, we partition the schema $\mathcal{D}$ into $K$ manageable batches $\{B_1, \dots, B_K\}$ where each batch consists of a subset of columns. For each batch $B_j$, the model identifies a deletion set of columns $\mathcal{C}_{del, j}$ and a preservation set of columns $\mathcal{C}_{keep, j}$. The pruned schema $\mathcal{D}_{pruned}$ is then formulated as:
% \begin{equation}
% \begin{aligned}
%     \mathcal{C}_{del, j} &= p_\theta(\cdot \mid q, P^*, B_j, F^{sl}_{del}) \\
%     \mathcal{C}_{keep, j} &= p_\theta(\cdot \mid q, P^*, B_j, F^{sl}_{sel}) \\
%     \mathcal{D}_{pruned} &= \bigcup_{j=1}^K \left( (B_j \setminus \mathcal{C}_{del, j}) \cup \mathcal{C}_{keep, j} \right)
% \end{aligned}
% \end{equation}
% where the union logic ensures that any ambiguous element is preserved unless it is both confidently rejected as noise and fails to be explicitly recognized as task-critical. 
% The detailed instruction templates $F^{sl}_{del}$ and $F^{sl}_{sel}$ are provided in \textcolor{red}{Appendix \ref{}}.

Exhaustive data exploration is computationally prohibitive, which motivates \TheName{} to introduce a preliminary filtration step. However, standard relevance filtering is prone to false negatives, often discarding foreign keys that lack both lexical and semantic alignment with the query terms. To address this issue, \TheName{} employs a \textbf{Dual-Pathway Pruning} strategy. Specifically, \TheName{} simultaneously executes a negative pass to remove obvious noise and a positive pass to identify relevant items, and subsequently fuses the results from both passes.
Let $\{B_1, \dots, B_K\}$ represent $K$ manageable batches partitioned from the schema $\mathcal{D}$, where each batch consists of a subset of columns. For each batch $B_j$, \TheName{} identifies a deletion set of columns $\mathcal{C}_{del, j} \subseteq B_j$ and a preservation set of columns $\mathcal{C}_{keep, j} \subseteq B_j$. The pruned schema $\mathcal{D}_{pruned}$ is then formulated as:

% \vspace{-3mm}
\begin{equation}\label{equ:schema_prune}
\begin{aligned}
    \mathcal{C}_{del, j} &= p_\theta(\cdot \mid q, P^*, B_j, F^{sl}_{del})\\
    \mathcal{C}_{keep, j} &= p_\theta(\cdot \mid q, P^*, B_j, F^{sl}_{sel}) \\
    \mathcal{D}_{pruned} &= \bigcup_{j=1}^K \left( (B_j \setminus \mathcal{C}_{del, j}) \cup \mathcal{C}_{keep, j} \right)
\end{aligned}
% \vspace{-1mm}
\end{equation}
where the union logic ensures that any ambiguous element is preserved unless it is both confidently rejected as noise and fails to be explicitly recognized as task-critical.
% The detailed instruction templates $F^{sl}_{del}$ and $F^{sl}_{sel}$ are provided in \textcolor{red}{Appendix \ref{}}.

% \vspace{-1.5mm}
\subsubsection{Hypothesis Verification via Agentic Exploration}
\label{sec:agentic_exploration}

To verify hypotheses in the logical plan over $\mathcal{D}_{pruned}$, \TheName{} employs \textbf{Agentic Exploration}, starting with \textbf{Semantic Linking}. In this step, \TheName{} hypothesizes the specific role of each table and column, such as serving as the primary filter.
This role mapping establishes the direction for the subsequent \textbf{Parallel Data Profiling}, in which \TheName{} assigns independent agents to each table, restricted to the subset of columns in $\mathcal{D}_{pruned}$. Unlike static profiling, these agents dynamically generate SQL queries grounded in the hypothesized roles to verify data patterns. Following this local exploration, the \textbf{Global Synthesis} phase enables \TheName{} to integrate the entire $\mathcal{D}_{pruned}$ along with their respective empirical observations, and to make the final decision on the sufficient subgraph $\mathcal{D}^{*}$.
This process can be formulated as follows:

% \vspace{-3mm}
\begin{equation}\label{equ:hypothesis_verification}
\begin{aligned}
    \mathcal{R} &= p_\theta(\cdot \mid q, \mathcal{D}_{pruned}, P^*, F^{sl}_{semantics}) \\
    \mathcal{E}_i &= \text{Exec}(p_\theta(\text{sql} \mid q, T_i, \mathcal{R}_i, F^{sl}_{exp}), \mathcal{D}_{pruned}) \\
    \mathcal{D}^* &= p_\theta(\cdot \mid q, \mathcal{D}_{pruned}, \{\mathcal{E}_i\}_{i=1}^m, F^{sl}_{final})
\end{aligned}
% \vspace{-1mm}
\end{equation}
where $\mathcal{R}$ denotes the functional role analysis, and $\mathcal{E}_i$ represents the empirical observations for each candidate table $T_i$. 
% The instruction templates $F^{sl}_{semantics}$, $F^{sl}_{exp}$, and $F^{sl}_{final}$ are detailed in \textcolor{red}{Appendix \ref{}}.

% \vspace{-1mm}
\subsection{Agentic Exploration for SQL Generation}

% While agentic exploration allows models to verify data assumptions, unguided exploration often proves unstable as models may perform superficial probes that overlook critical edge cases. Thus, we propose an Agentic Framework that disciplines the generation process through deterministic guidance retrieval and fidelity-oriented profiling.

% Synthesizing correct SQL requires bridging the gap between abstract user intent and rigid database schemas. 
% Therefore, we implement $T^3D$ through the Agentic Profiling framework, which interleaves reasoning and exploration, allowing the model to form and validate hypotheses rather than rely on blind guessing. To ensure stable performance, the system first retrieves relevant guidance that prevents meaningless or superficial probing.

To translate abstract user intent into executable SQL under rigid schema constraints, \TheName{} instantiates \emph{agentic exploration within the SQL generation stage} through the H-V loop.
Specifically, \TheName{} interleaves reasoning and exploration to actively synthesize hypotheses and  validate them through evidence-driven query execution (Figure \ref{fig:framwork} (Bottom)).
To ensure meaningful exploration, \TheName{} first retrieves question-relevant guidance that constrains the search space and prevents superficial or degenerate probing.

% \vspace*{-1.5mm}
\subsubsection{Guidance via Deterministic Retrieval}
\label{sec:derterminitic_retrieval}

Exploration is only effective when the agent knows \textit{what} to verify. Without explicit guidance, models are prone to overlooking critical constraints such as integer division truncation or case-sensitivity. To address this limitation, \TheName{} introduces a \textbf{Deterministic Retrieval} mechanism that leverages the logical plan (Section~\ref{sec: logical plan}) to retrieve query-dependent exploration guidance. 
Specifically, \TheName{} infers potential SQL realization paths for each logical step, such as implementing ``ranking'' using the \texttt{RANK()} function over the \texttt{score} column.
Based on these realizations, \TheName{} applies a rule engine to extract and match operational keywords against a predefined library, such that detecting a \texttt{RANK} keyword activates specific directives, including inspecting \texttt{NULL} values in the corresponding column.
This process is formulated as:

% \vspace{-3mm}
\begin{equation}\label{equ:guidance}
\begin{aligned}
    \mathcal{W} &= \text{ExtractKeywords}\!\left(p_\theta(\cdot \mid q, P^*, \mathcal{D}^*, F^{sql}_{kw})\right) \\
    \mathcal{G} &= \{ m \in \mathcal{M} \mid \text{Match}(\mathcal{W}, m) = \text{True} \}
\end{aligned}
% \vspace{-1mm}
\end{equation}
where $\mathcal{W}$ denotes the set of operational keywords, $F^{sql}_{kw}$ is the instruction template, and $\mathcal{M}$ represents a library of SQL best practices. 
Since SQL operations constitute a finite and closed set, \TheName{} adopts deterministic keyword mapping to achieve higher reliability than dense retrieval, which often introduces irrelevant noise.
As a result, \TheName{} transforms the logical plan from a passive blueprint into an executable verification guide, ensuring that exploration remains focused on relevant technical risks.

\begin{table*}[t]
% \scriptsize
\centering
\setlength{\tabcolsep}{8.5pt}
\caption{Schema linking performance evaluating Strict Recall Rate (SRR), Non-Strict Recall (NSR), Non-Strict Precision (NSP) and Non-Strict F1-Score (NSF). Evaluation settings include BIRD Subset (DeepSeek-V3.2), BIRD Full Set (Qwen3-32B), and Spider 2.0-Snow Subset (GPT-4.1). All baseline results in this table are locally reproduced.}
\label{tab:schema_linking_main}
% \resizebox{\textwidth}{!}{
\begin{tabular}{l cccc cccc cccc}
\toprule
\multirow{2}{*}{\textbf{Method}} & \multicolumn{4}{c}{\textbf{BIRD ($N=147$)}} & \multicolumn{4}{c}{\textbf{BIRD ($N=1534$)}} & \multicolumn{4}{c}{\textbf{Spi.-Snow ($N=120$)}} \\
\cmidrule(lr){2-5} \cmidrule(lr){6-9} \cmidrule(lr){10-13}
& \textbf{SRR} & NSR & NSP & NSF & \textbf{SRR} & NSR & NSP & NSF & \textbf{SRR} & NSR & NSP & NSF \\
\midrule
\rowcolor{green!5}
& \multicolumn{12}{c}{\textbf{Training-based Methods}} \\
\midrule
CodeS & 59.86 & 86.53 & 31.41 & 44.84 & 67.08 & 88.06 & 30.64 & 43.90 & \ \ 8.30 & 34.15 & 20.16 & 25.35 \\
RESDSQL & 46.24 & 74.59 & 27.70 & 39.23 & 49.61 & 76.24 & 27.48 & 38.82 & 23.30 & 55.64 & 27.42 & 36.74 \\
\midrule
\rowcolor{green!5}
& \multicolumn{12}{c}{\textbf{Training-free Methods}} \\
\rowcolor{green!5}
& \multicolumn{4}{c}{\textit{Backbone: DeepSeek-V3.2}} & \multicolumn{4}{c}{\textit{Backbone: Qwen3-32B}} & \multicolumn{4}{c}{\textit{Backbone: GPT-4.1}} \\
\midrule
TA-SQL & 82.99 & 93.47 & \textbf{80.19} & \textbf{84.96} & 75.55 & 90.71 & \textbf{81.71} & \textbf{84.35} & 67.08 & 88.06 & 30.64 & 45.46 \\
RSL-SQL & 91.84 & 97.74 & 37.83 & 51.30 & 85.92 & 95.75 & 47.90 & 60.43 & 80.83 & 88.19 & 25.43 & 35.24 \\
ReFoRCE & 34.01 & 67.50 & 76.07 & 67.81 & 19.30 & 51.29 & 68.05 & 55.02 & 35.00 & 69.58 & 53.73 & 54.33 \\
DSR-SQL & 89.12 & 96.78 & 25.48 & 38.41 & 86.70 & \textbf{96.23} & 23.65 & 35.77 & 53.33 & 82.02 & \textbf{57.05} & \textbf{64.10} \\
AutoLink & 93.88 & 96.91 & 10.29 & 17.50 & 39.50 & 74.39 & 16.27 & 24.86& 85.83 & 94.55 & \ \ 8.11 & 14.18 \\
\midrule
\TheName{} & \textbf{97.28} & \textbf{99.35} & 39.02 & 52.78 & \textbf{87.68} & 96.15 & 51.95 & 64.10 & \textbf{88.33} & \textbf{97.08} & 36.52 & 46.96 \\
\bottomrule
\end{tabular}
% }
% \vspace*{-2mm}
\end{table*}

% \vspace*{-1.5mm}
\subsubsection{Grounding via Agentic Exploration}
\label{sec:sql_exploration}

Guided by $\mathcal{G}$, \TheName{} initiates an agentic process to resolve implementation ambiguities within $\mathcal{D}^*$ by transitioning over the action space $\mathcal{A}$ through a sequence of states $H_t$, where each new state is determined based on the prior history $H_{<t}$ and the accumulated observations $O_{<t}$:

% \vspace{-3mm}
\begin{equation}
H_t = p_\theta(\cdot \mid \mathcal{A}, \mathcal{G}, H_{<t}, O_{<t}).
\end{equation}
% \vspace{-5mm}

\paragraph{Profiling}
To resolve uncertainties regarding value distributions and data formats, \TheName{} autonomously generates exploratory queries $Q_{exp, t}$ at iteration $t$. To prevent large result sets from flooding the context window, \TheName{} automatically compresses extensive outputs into a statistical summary $O_t = \sigma(\text{Exec}(Q_{exp, t}, \mathcal{D}^*))$, providing a compact yet comprehensive view of the data landscape.

% \vspace*{-1.5mm}
\paragraph{Consolidation}
As multi-step exploration progresses, \TheName{} performs periodic state compression to prevent context saturation and maintain logical coherence as evidence accumulates. Once sufficient observations are collected, \TheName{} synthesizes a consolidated status update that aligns observations with the overall logical plan and refines its understanding of schema semantics.

% \vspace*{-1.5mm}
\paragraph{SQL Synthesis}
After all implementation details are grounded in empirical evidence, \TheName{} synthesizes a candidate SQL query $S$ that maps the refined reasoning state onto the physical database syntax. The synthesized query is immediately validated against the database environment to ensure execution feasibility. If the execution $\text{Exec}(S, \mathcal{D}^*)$ returns an error, \TheName{} leverages the resulting feedback to drive self-correction in subsequent iterations.

% \vspace*{-1.5mm}
\paragraph{Confirmation}
Following a successful execution, \TheName{} enters a terminal state to perform secondary verification of semantic fidelity. Specifically, \TheName{} evaluates whether the SQL logic aligns with the user requirements, the accumulated observations, and the latest logical plan. The output is finalized only after \TheName{} confirms these conditions, thereby preventing logical drift during the multi-step generation process.

\section{Experiments}
Our evaluation focuses on the effectiveness of schema linking in reducing the search space while preserving relevant columns, and the downstream execution accuracy of the generated SQL queries.

% \vspace{-1.5mm}
\subsection{Datasets}
% We conduct experiments on BIRD~\cite{li2023can} and Spider 2.0~\cite{lei2024spider}, which capture the complexity of real-world enterprise databases.

% % \vspace*{-2mm}
\noindent\textbf{BIRD-Dev.} The development set of BIRD~\cite{li2023can}, which is a large-scale cross-domain dataset consists of 1,534 queries across 11 databases. It requires the model to resolve ambiguities in fine-grained schema details, such as similarly named columns or dirty database values.

% % \vspace*{-2mm}
\noindent\textbf{Spider 2.0-Snow.} Spider 2.0~\cite{lei2024spider} represents the frontier of enterprise-level Text-to-SQL tasks. Its primary challenge lies in its massive schema size, featuring an average of over 800 columns per database and highly complex SQL structures. We focus on the Spider 2.0-Snow sub-task, which requires generating SQL queries in the Snowflake dialect, a standard for modern cloud data warehouses.\footnote{\label{fn:spider2_version}The Spider 2.0 evaluation script has been updated since its initial release. The original script adopts a relaxed matching protocol, while the updated version applies a more lenient evaluation criterion. Results obtained under different script versions are not directly comparable. In Table~\ref{tab:sql_generation_results_spider}, we mark results with $\dagger$ (original script) and $\ddagger$ (updated script) accordingly. Note that DSR-SQL~\cite{hao2025text} also reports results under both versions in their appendix (35.28\% vs.\ 52.83\%).}

% \vspace{-1.5mm}
\subsection{Experiment 1: Schema Linking Performance}

\subsubsection{Metrics}
We evaluate schema linking as a retrieval task, aiming to identify the minimal sufficient subgraph required for the query. We follow prior work~\cite{cao2024rsl} to report the following metrics:
\begin{itemize}
[leftmargin=*,itemsep=0pt,parsep=0.1em,topsep=0.1em,partopsep=0.1em]
\item\textbf{Strict Recall Rate (SRR)} measures the percentage of queries where the retrieved schema fully covers the ground truth. We consider SRR the most important metric, as it directly determines whether downstream SQL generation is possible to succeed.
\item{Non-Strict Recall (NSR)} computes the proportion of required columns retrieved, even if the full subgraph is incomplete.
\item{Non-Strict Precision (NSP)} measures the ratio of relevant columns in the retrieved set, reflecting the model's ability to filter noise.
\item{Non-Strict F1-Score (NSF)} is the harmonic mean of NSR and NSP, offering a holistic view of retrieval quality.
\end{itemize}

% \vspace*{-1.5mm}
\subsubsection{Experimental Setup}
To ensure a rigorous evaluation, we define specific testing splits and model backbones for each dataset. 
For BIRD-Dev, we use a widely adopted subset of 147 cases~\cite{talaei2024chess,lialpha} to evaluate the performance of DeepSeek-V3.2~\cite{liu2024deepseek} as the backbone, allowing us to test the model's capability within a controlled range and replicate more baselines. For full-set evaluation, we use Qwen3-32B~\cite{yang2025qwen3}.
For Spider 2.0-Snow, we utilize GPT-4.1~\cite{GPT4} on the 120-case subset with official golden SQLs\footnote{Full execution ground truth is available, but golden SQLs cover only 120 cases.}, selected for its extended context window, which ensures both baseline and our method are not limited by context window size when processing Spider 2.0's larger schema.
Ground truth columns are obtained by using both GPT-4.1 and DeepSeek-V3.2 to extract referenced columns from the golden SQLs, followed by manual verification.
The instruction templates $F^{sl}_{plan}$ and $F^{sl}_{agg}$ in Equ.~\ref{equ:plan_agg}, $F^{sl}_{del}$ and $F^{sl}_{sel}$ in Equ.~\ref{equ:schema_prune}, $F^{sl}_{semantics}$, $F^{sl}_{exp}$, and $F^{sl}_{final}$ in Equ.~\ref{equ:hypothesis_verification} are provided in Appendix \ref{app:prompts}.

% \vspace*{-1.5mm}
\subsubsection{Baselines}
We compare our approach against comprehensive baselines, including CodeS~\cite{li2024codes} and RESDSQL~\cite{li2023resdsql}, which are training-base methods, and TA-SQL~\cite{qu2024before}, RSL-SQL~\cite{cao2024rsl}, ReFoRCE~\cite{deng2025reforce}, DSR-SQL~\cite{hao2025text}, and AutoLink~\cite{wang2025autolink}, which are training-free methods.

% \vspace*{-1.5mm}
\subsubsection{Results on Schema Linking}
Table \ref{tab:schema_linking_main} presents the schema linking performance across two benchmarks and three experimental settings. \TheName{} achieves the highest SRR across all settings, recording 97.28\% on the BIRD subset, 87.68\% on the BIRD full set, and 88.33\% on Spider 2.0-Snow.
We observe that existing training-based methods perform poorly on Spider 2.0, with significant drops in SRR. Among training-free methods, although TA-SQL, ReFoRCE, and DSR-SQL (on Spider 2.0) achieve higher NSP, their SRR is much lower than ours (e.g., ReFoRCE achieves 35.00\% on Spider 2.0), rendering them unsuitable for downstream SQL generation, where missing columns can lead to critical errors.
While AutoLink achieves an SRR close to ours on the BIRD subset and Spider 2.0, its NSP is noticeably lower, highlighting the superiority of \TheName{} in maintaining precision while maximizing recall.
Furthermore, \TheName{} better adapts to weaker models, whereas AutoLink shows a significant drop in SRR on the BIRD full set (with Qwen3-32B).

% \vspace*{-1.5mm}
\subsubsection{Schema Linking Details}
Logical Planning: We sample $N=2$ paths at temperature 0.8 using only the query $q$ as input, then aggregate them at temperature 0.2.
Dual-Pathway Pruning: Columns are batched by table to maintain semantic integrity, with each batch constrained to 8--12k tokens. To minimize API overhead, we implement \textit{Schema Merging}: tables with identical columns are consolidated into a single prompt entry that displays one representative schema alongside a list of all matching table names. 

\begin{table}[t]
\centering
\caption{Execution accuracy on BIRD-Dev ($N=1534$) across difficulty levels. Results marked with $^*$ are locally reproduced, while others are collected from existing literature.}
\label{tab:sql_generation_results}
\setlength{\tabcolsep}{3.1pt}
% \resizebox{\columnwidth}{!}{
\begin{tabular}{llcccc}
\toprule
\multirow{2}{*}{\textbf{Method}} & \multirow{2}{*}{\textbf{LLM}} & \textbf{Sim.} & \textbf{Mod.} & \textbf{Chall.} & \textbf{Total} \\
& & (925) & (464) & (145) & (1534) \\
\midrule
C3-SQL & GPT-4 & 58.9 & 38.5 & 31.9 & 50.2 \\
% DIN-SQL & GPT-4 & & & 50.7 \\
DAIL-SQL & GPT-4 & 62.5 & 43.2 & 37.5 & 54.3 \\
TA-SQL & GPT-4 & 63.1 & 48.6 & 36.1 & 56.2 \\
MAG-SQL & GPT-4 & 65.9 & 46.2 & 41.0 & 57.6 \\
SuperSQL & GPT-4 & 66.9 & 46.5 & 43.8 & 58.5 \\
MAC-SQL & GPT-4 & 65.7 & 52.7 & 40.3 & 59.4 \\
MCS-SQL & GPT-4 & 70.4 & 53.1 & 51.4 & 63.4 \\
Contextual-SQL$^*$ & GPT-4o & 72.9 & 60.3 & 52.4 & 67.1 \\
RSL-SQL & GPT-4o & 74.4 & 57.1 & 53.8 & 67.2 \\
CHESS & Gemini-1.5-Pro & - & - & - & 68.3 \\
DSR-SQL & DeepSeek-V3.1 & 72.7 & 61.2 & \textbf{63.5} & 68.3 \\
AutoLink & Gemini-1.5-Pro & - & - & - & 68.7 \\
OpenSearch-SQL & GPT-4o & - & - & - & 69.3 \\
\midrule
\TheName{} & GPT-4o & \textbf{75.9} & \textbf{64.4} & 57.2 & \textbf{70.7} \\
\bottomrule
\end{tabular}
% }
\end{table}

% \vspace*{-1.5mm}
\subsection{Experiment 2: SQL Generation Performance}

\subsubsection{Metrics}
We use \textbf{Execution Accuracy (EX)} as the evaluation metric, measuring the percentage of generated queries that return the exact result set as the ground truth\footnote{Compared to BIRD, Spider 2.0 adopts a relaxed evaluation protocol due to output format ambiguity in complex queries, tolerating irrelevant columns in the result set.}. In addition, following prior work~\cite{deng2025reforce}, we also report \textbf{Pass@8} on Spider 2.0-Snow, where a question is considered solved if at least one correct SQL appears among eight candidates. Unless specified otherwise, we use the results reported in the official papers of the respective baselines.

% \vspace{-0.5mm}
\subsubsection{Baselines}
We compare \TheName{} against open-source methods or those that have been reproduced by other works. For \textbf{BIRD-Dev}, we benchmark against C3-SQL~\cite{dong2023c3}, DAIL-SQL~\cite{gao2023text}, TA-SQL~\cite{qu2024before}, MAG-SQL~\cite{xie2024mag}, SuperSQL~\cite{li2024dawn}, MAC-SQL~\cite{wang2025mac}, Contextual-SQL~\cite{agrawal2025text2sql}, MCS-SQL~\cite{lee2025mcs}, RSL-SQL~\cite{cao2024rsl}, CHESS~\cite{talaei2024chess}, DSR-SQL~\cite{hao2025text}, AutoLink~\cite{wang2025autolink}, and OpenSearch-SQL~\cite{xie2025opensearchsqlenhancingtexttosqldynamic}. For \textbf{Spider 2.0-Snow}, we compare against Spider-Agent~\cite{lei2024spider}, ReFoRCE~\cite{deng2025reforce}, and DSR-SQL~\cite{hao2025text}.

\begin{table}[t]
\centering
\caption{Main results on Spider 2.0-Snow ($N=547$), evaluating execution accuracy (EX) and solution coverage (Pass@8). Baseline results are as reported in the literature. $\dagger$ denotes results evaluated under the original evaluation script; $\ddagger$ denotes results evaluated under the updated evaluation script (see Footnote~\ref{fn:spider2_version}).}
\label{tab:sql_generation_results_spider}
\setlength{\tabcolsep}{11pt}
\begin{tabular}{llcc}
\toprule
\textbf{Method} & \textbf{LLM} & \textbf{EX} & \textbf{Pass@8} \\
\midrule
% CHESS & GPT-4o & 1.28 \\
% Spider-Agent & GPT-4o & 10.05 \\
Spider-Agent$^\dagger$ & DeepSeek-R1 & 10.79 & - \\
Spider-Agent$^\dagger$ & o1-preview & 23.77 & - \\
% ReFORCE & GPT-4o & 20.84 \\
ReFORCE$^\dagger$ & DeepSeek-R1 & 29.25 & - \\
ReFORCE$^\dagger$ & o4-mini & 29.80 & 31.99 \\
ReFORCE$^\dagger$ & o3 & 35.83 & 39.85 \\
DSR-SQL$^\dagger$ & DeepSeek-R1 & 35.28 & - \\
DSR-SQL$^\ddagger$ & DeepSeek-R1 & 52.83 & - \\
% AutoLink & DeepSeek-R1 & \\
\midrule
\TheName{}$^\ddagger$ & DeepSeek-R1 & \textbf{53.03} & \textbf{68.44} \\
\bottomrule
\end{tabular}
\end{table}

\begin{table*}[t]
\centering
\caption{Cross-Model Comparison on the Spider 2.0-Snow subset ($N=120$) with oracle schema under three settings: \textbf{w/ Exploration} (ours), \textbf{w/o Exploration} (pure SQL generation with execution-guided refinement), and \textbf{Oracle Exploration} (implementation details derived from Ground Truth SQL), evaluated using EX, EX@8 (average EX score across 8 results), and Pass@8 metrics.}
\label{tab:exploration}
\setlength{\tabcolsep}{7.3pt}
\begin{tabular}{l cccc cccc cccc}
\toprule
\multirow{2}{*}{\textbf{Base Model}} & \multicolumn{3}{c}{\textbf{w/ Exploration (Ours)}} & \multicolumn{3}{c}{\textbf{w/o Exploration}} & \multicolumn{3}{c}{\textbf{Oracle Exploration}} \\
& \textbf{EX} & \textbf{EX@8} & \textbf{Pass@8} & \textbf{EX} & \textbf{EX@8} & \textbf{Pass@8} & \textbf{EX} & \textbf{EX@8} & \textbf{Pass@8} \\
\midrule
GPT-4o & 41.67 & 29.79 & 55.83 & $36.67_{{\color{my_red}\downarrow\ 5.00}}$ & $24.17_{\color{my_red}{\downarrow\ 5.62}}$ & $51.67_{\color{my_red}{\downarrow\ 4.16}}$ & $45.83_{\color{my_green}{\uparrow4.16}}$ & $35.73_{{\color{my_green}\uparrow5.94}}$ & $62.50_{{\color{my_green}\uparrow6.67}}$ \\
Kimi-k2-instruct & 48.33 & 41.46 & 70.00 & $38.33_{{\color{my_red}\downarrow10.00}}$ & $32.50_{{\color{my_red}\downarrow\ 8.96}}$ & $52.50_{{\color{my_red}\downarrow17.50}}$ & $57.50_{{\color{my_green}\uparrow9.17}}$ & $46.98_{{\color{my_green}\uparrow5.52}}$ & $65.83_{{\color{my_red}\downarrow4.17}}$ \\
GPT-5 & 55.00 & 47.92 & 73.33 & $40.83_{{\color{my_red}\downarrow14.17}}$ & $31.77_{{\color{my_red}\downarrow16.15}}$ & $55.00_{{\color{my_red}\downarrow18.33}}$ & $56.67_{{\color{my_green}\uparrow1.67}}$ & $52.19_{{\color{my_green}\uparrow4.27}}$ & $69.17_{{\color{my_red}\downarrow4.16}}$ \\
DeepSeek-R1 & 57.50 & 48.75 & 79.17 & $51.67_{{\color{my_red}\downarrow\ 5.83}}$ & $37.50_{{\color{my_red}\downarrow11.25}}$ & $69.17_{{\color{my_red}\downarrow10.00}}$ & $58.33_{\color{my_green}\uparrow0.83}$ & $55.83_{\color{my_green}\uparrow7.08}$ & $78.33_{\color{my_red}\downarrow0.84}$ \\
DeepSeek-V3.2 & 57.50 & 52.71 & 79.17 & $39.17_{{\color{my_red}\downarrow18.33}}$ & $32.92_{{\color{my_red}\downarrow19.79}}$ & $52.50_{{\color{my_red}\downarrow26.67}}$ & $55.83_{{\color{my_red}\downarrow1.67}}$ & $48.13_{{\color{my_red}\downarrow4.58}}$ & $70.83_{{\color{my_red}\downarrow8.34}}$ \\
GPT-4.1 & 62.50 & 47.19 & 74.17 & $44.17_{{\color{my_red}\downarrow18.33}}$ & $35.21_{{\color{my_red}\downarrow11.98}}$ & $59.17_{{\color{my_red}\downarrow15.00}}$ & $63.33_{{\color{my_green}\uparrow0.83}}$ & $52.19_{{\color{my_green}\uparrow5.00}}$ & $75.00_{{\color{my_green}\uparrow0.83}}$ \\
Gemini-3-Pro & 66.67 & \textbf{63.33} & 78.33 & $57.50_{{\color{my_red}\downarrow\ 9.17}}$ & $41.46_{{\color{my_red}\downarrow21.87}}$ & $62.50_{{\color{my_red}\downarrow15.83}}$ & $70.00_{{\color{my_green}\uparrow3.33}}$ & $60.94_{{\color{my_red}\downarrow2.39}}$ & $78.33_{{\color{my_red}\downarrow0.00}}$ \\
Claude-4.5-Sonnet & \textbf{67.50} & 61.67 & \textbf{80.83} & $50.83_{{\color{my_red}\downarrow16.67}}$ & $48.44_{{\color{my_red}\downarrow13.23}}$ & $64.17_{{\color{my_red}\downarrow16.66}}$ & $69.17_{{\color{my_green}\uparrow1.67}}$ & $67.92_{{\color{my_green}\uparrow6.25}}$ & $79.17_{{\color{my_red}\downarrow1.66}}$ \\
\bottomrule
\end{tabular}
\end{table*}

% \vspace{-0.5mm}
\subsubsection{Result Analysis}
As presented in Tables \ref{tab:sql_generation_results} and \ref{tab:sql_generation_results_spider}, \TheName{} consistently outperforms mainstream approaches.
On \textbf{BIRD-Dev}, \TheName{} achieves \textbf{70.7\%} accuracy, surpassing retrieval-augmented methods like DAIL-SQL ($54.3\%_{\downarrow16.4}$) and RSL-SQL ($67.2\%_{\downarrow3.5}$). Despite slightly trailing DSR-SQL in the \textit{Challenging} subset, \TheName{} dominates the \textit{Simple} and \textit{Moderate} categories, demonstrating superior robustness against semantic ambiguity.
This advantage extends to \textbf{Spider 2.0-Snow}. As shown in Table~\ref{tab:sql_generation_results_spider}, \TheName{} achieves \textbf{53.03\%} EX with DeepSeek-R1 under the updated evaluation script. While direct numerical comparison with prior methods is complicated by the evaluation script change (Footnote~\ref{fn:spider2_version}), for reference, DSR-SQL reports 35.28\% (original) and 52.83\% (updated) under the two versions respectively, suggesting our method remains competitive or superior under the same evaluation protocol.
Notably, \TheName{} attains a Pass@8 of \textbf{68.44\%}, indicating that our guided exploration effectively maximizes solution coverage. 
The comparison with DSR-SQL further highlights \TheName{}'s strength: while DSR-SQL incorporates data profiling in its SQL generation process, it lacks agentic exploration in the schema linking phase and guidance for exploration. 

% \vspace{-0.5mm}
\subsubsection{SQL Generation Details}
\label{sec:sql_details}
We implement all agentic components in a training-free manner, with detailed instructions provided in Appendix \ref{app:prompts}. 
Guidance Retrieval: We map inferred natural language paths to the library $\mathcal{M}$ 
(provided in Appendix \ref{appx:library}) 
using keyword matching, which achieved \textbf{$>95\%$} recall in pilot tests on a BIRD-train subset.
Data Profiling: The profiling function $\sigma$ compresses result sets exceeding 30 rows into the top 10 rows plus statistics (row count, cardinality, types, and NULL ratios). For each question, we enforce a limit of 40 actions and $56k$ tokens; a mandatory SQL synthesis action is triggered at 38 actions or 52k tokens. For Spider 2.0-Snow cases with external knowledge (up to $30k$ tokens), we pre-filter the documentation for query-relevant snippets using the tested model. 
% (Appx. \ref{app:prompts}).
Context Management: During consolidation, we prune the interaction history to retain only exploratory queries, execution results, and the latest consolidated plan/understanding.
Inference: We sample 8 candidates per task. For BIRD, we select the final answer using a reward model following~\cite{agrawal2025text2sql}. For Spider 2.0-Snow, we use result-based majority voting with model selection
(Appx. \ref{app:prompts})
as a tie-breaker.
The instruction $F^{sql}_{kw}$ for generating SQL realization paths, and the definition of the Action space $\mathcal{A}$ are also provided in Appendix \ref{app:prompts}.

% \vspace*{-1.5mm}
\subsection{Analysis}
In this section, we present a series of experiments analyzing the impact of autonomous exploration, the scalability of test-time compute, and the contributions of core components in both the schema linking and SQL generation stages. We also present a case study in Appendix~\ref{appx:case_study}.
All experiments are conducted on the Spider 2.0-Snow subset, which includes official golden SQLs.
We prioritize this benchmark due to its higher complexity, which provides more discriminative metrics, effectively highlighting the performance boundaries of different strategies.
For SQL generation-related investigations (Sections \ref{analysis:exploration} and \ref{analysis:passk}), we consistently use the \textit{oracle schema} setting, where models are provided with ground truth schema information. This approach isolates the impact of schema linking errors, ensuring that performance differences are attributed solely to SQL generation. We also introduce the \textbf{EX@8} metric, which calculates the average EX score across 8 results.
% , providing a more comprehensive view of performance.

\subsubsection{Effect of Exploration}
\label{analysis:exploration}
To quantify the contribution of agentic exploration, we compare three settings: \textbf{w/ Exploration} (ours), \textbf{w/o Exploration} (pure SQL generation with execution-guided refinement), and \textbf{Oracle Exploration}\footnote{\textbf{Oracle Exploration} injects implementation details derived from Ground Truth SQL via GPT-5, including parsing strategies and implicit data handling steps that are not explicitly mentioned in the query or schema, such as how to join multiple tables.}. 
Table \ref{tab:exploration} presents the results, revealing three critical insights:

% \vspace*{-1.5mm}
\paragraph{Exploration as a Performance Multiplier} 
Exploration significantly improves performance across all metrics. For instance, DeepSeek-V3.2’s EX score increases from 39.17\% to 57.50\% ($\uparrow$18.33\%). This shows that reasoning alone cannot fully resolve the gap between abstract logic and physical data states—empirical verification is essential for generating accurate enterprise-level SQL queries.

% \vspace*{-1.5mm}
\paragraph{The ``Rich Get Richer'' Effect} 
We observe that exploration gains correlate with base model capability. Weaker models like GPT-4o show modest improvements ($\uparrow$5.00\% in EX), while stronger models such as DeepSeek-V3.2 and Gemini-3-Pro experience much larger gains (e.g., DeepSeek-V3.2 $\uparrow$18.33\% in EX). This suggests that stronger models possess superior cognitive agility, allowing them to better formulate hypotheses, interpret exploration feedback, and synthesize this information into more accurate SQL logic.

% \vspace*{-1.5mm}
\paragraph{Autonomy vs. Oracle Guidance}
Comparing autonomous exploration (w/ Exploration) with Oracle Exploration reveals a capability-dependent trade-off. For weaker models like GPT-4o, Oracle guidance consistently improves all metrics (e.g., Pass@8 $\uparrow$6.67\%), suggesting that models with limited reasoning capacity benefit more from shortcuts.
However, for stronger models, while Oracle improves EX, it often hurts Pass@8 (e.g., DeepSeek-V3.2 $\downarrow$8.34\%). This indicates that while Oracle narrows the search space, autonomous exploration provides a broader understanding of the data. For DeepSeek-V3.2 and Gemini-3-Pro, self-exploration even leads to higher EX@8 scores than Oracle, indicating that stronger models more consistently benefit from exploration.

\begin{figure}[t]
    \centering
    \includegraphics[width=0.9\linewidth]{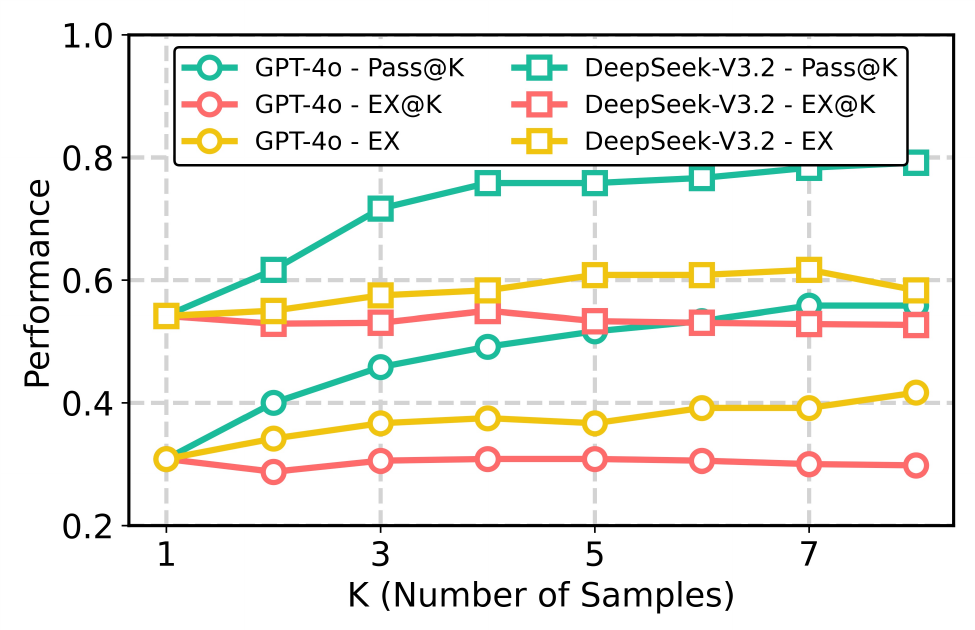}
    % \vspace{-3mm}
    \caption{Performance scaling of GPT-4o and DeepSeek-V3.2 on Spider 2.0-Snow subset ($N=120$) using the oracle schema.}
    \label{fig:pass_k_analysis}
\end{figure}

% \vspace{-1.5mm}
\subsubsection{Test-Time Compute Scaling}
\label{analysis:passk}
To investigate the relationship between inference budget and performance scaling, we conduct a scalability analysis by performing $N=8$ independent generation runs using the oracle schema. 
Figure \ref{fig:pass_k_analysis} reveals three key insights:

% \vspace*{-1.5mm}
\paragraph{Expanding Latent Potential} The growth of Pass@k confirms that scaling test-time compute effectively expands the solution search space, uncovering the models' latent reasoning potential. Notably, with a budget of $k=8$, the coverage of GPT-4o ($Pass@8=55.8\%$) surpasses the average single-pass performance of DeepSeek-V3.2 ($EX@k=52.7\%$), suggesting that test-time compute scaling can effectively compensate for intrinsic model capabilities.

% \vspace*{-1.5mm}
\paragraph{Performance Consistency \& The Selection Gap} While Pass@k rises significantly, EX@k remains stable, suggesting high fidelity per trace rather than relying on brute-force sampling to rescue low-quality attempts. 
However, a substantial gap persists between the potential maximum (Pass@k) and the voting result (EX). For instance, DeepSeek-V3.2 achieves 79.17 Pass@8, yet voting captures only 57.50. This highlights that while the generative capability is sufficient, \textit{answer selection} remains a bottleneck.
In this work, we employ multiple sampling to mitigate generation variance rather than as a primary source of performance improvement, leaving the optimization of answer selection to future research.

\begin{table}[t]
\centering
\caption{Ablation study on schema linking components. The evaluation is stratified into two stages: the \textbf{Pruning} Stage (top), measuring the quality of the intermediate candidate columns before verification, and the \textbf{Verification Stage} (bottom), assessing the final output.
$\bar{C}$ denotes the average number of columns retained per query (3,430 before pruning).}
\label{tab:ablation_sl}
\setlength{\tabcolsep}{2pt}
\begin{tabular}{l ccccc}
\toprule
\multirow{2}{*}{\textbf{Method}} & \multicolumn{5}{c}{\textbf{Spider 2.0-Snow ($N=120$)}} \\
& \textbf{SRR} & NSR & NSP & NSF & $\bar{C}$ \\
\midrule
\rowcolor{green!5}
\multicolumn{6}{c}{\textbf{Workflow: Planning \& Pruning (w/o Verification)}} \\
\midrule
\TheName{} & \textbf{97.5} & \textbf{99.5} & 11.6 & 19.3 & 383 \\
\multicolumn{5}{l}{\textbf{\textit{Logical Planning}}} \\
w/o Planning & $97.2_{\color{my_red}\downarrow\ 0.3}$ & $99.2_{\color{my_red}\downarrow0.3}$ & $11.7_{\color{my_green}\uparrow\ 0.1}$ & $19.5_{\color{my_green}\uparrow\ 0.2}$ & 375 \\
\multicolumn{5}{l}{\textbf{\textit{Dual-Pathway Pruning}}} \\
Only Deletion & $80.8_{\color{my_red}\downarrow16.7}$ & $95.4_{\color{my_red}\downarrow4.1}$ & $\textbf{29.0}_{\color{my_green}\uparrow17.4}$ & $\textbf{39.4}_{\color{my_green}\uparrow20.1}$ & 217 \\
Only Selection & $93.3_{\color{my_red}\downarrow\ 4.2}$ & $98.5_{\color{my_red}\downarrow1.0}$ & $27.6_{\color{my_green}\uparrow16.0}$ & $37.6_{\color{my_green}\uparrow18.3}$ & 305 \\
\midrule
\rowcolor{green!5}
\multicolumn{6}{c}{\textbf{Workflow: Planning \& Pruning \& Verification}} \\
\midrule
\TheName{} & \textbf{77.5} & \textbf{94.9} & 31.7 & 43.0 & 57 \\
\multicolumn{5}{l}{\textbf{\textit{Logical Planning}}} \\
w/o Planning & $72.5_{\color{my_red}\downarrow\ 5.0}$ & $94.6_{\color{my_red}\downarrow\ 0.3}$ & $37.2_{\color{my_green}\uparrow\ 5.5}$ & $49.5_{\color{my_green}\uparrow\ 6.5}$ & 33 \\
\multicolumn{5}{l}{\textbf{\textit{Agentic Verification}}} \\
w/o Semantic Linking & $63.3_{\color{my_red}\downarrow14.2}$ & $91.3_{\color{my_red}\downarrow\ 3.6}$ & $37.2_{\color{my_green}\uparrow\ 5.5}$ & $48.2_{\color{my_green}\uparrow\
 5.2}$ & 30 \\
w/o Data Profiling & $68.3_{\color{my_red}\downarrow\ 9.2}$ & $90.8_{\color{my_red}\downarrow\ 4.1}$ & $41.4_{\color{my_green}\uparrow\ 9.7}$ & $53.4_{\color{my_green}\uparrow10.4}$ & 25 \\
w/o Global Synthesis & $73.3_{\color{my_red}\downarrow\ 4.2}$ & $94.4_{\color{my_red}\downarrow\ 0.5}$ & $38.7_{\color{my_green}\uparrow\ 7.0}$ & $51.0_{\color{my_green}\uparrow\ 8.0}$ & 45 \\
w/o All & $55.8_{\color{my_red}\downarrow21.7}$ & $82.7_{\color{my_red}\downarrow12.2}$ & $\textbf{63.5}_{\color{my_green}\uparrow31.8}$ & $\textbf{69.8}_{\color{my_green}\uparrow26.8}$ & 12 \\
\bottomrule
\end{tabular}
\end{table}

% \vspace{-1.5mm}
\subsubsection{Ablation Study on Schema Linking Components}
\label{analysis:ablation_sl}
We perform an ablation study to evaluate the contributions of key components in schema linking. Results are shown in Table \ref{tab:ablation_sl}.

% \vspace*{-1.5mm}
\paragraph{The Recall-Precision Trade-off.} 
We observe that our full method achieves lower precision (NSP) compared to baselines. We argue this behavior is accepted, as enterprise queries often allow for multiple valid solution paths. For instance, returning either \texttt{id} or \texttt{name} is acceptable for entity identification requirements. While ground truth SQL may select one, a robust schema linker must preserve all candidates to avoid irreversible pruning errors. As a result, we trade reduced precision for higher SRR, as missing a single critical column renders correct SQL generation impossible.

% \vspace*{-1.5mm}
\paragraph{Logical Planning.} 
Removing \textit{Logical Planning} minimally impacts pruning stage SRR ($\downarrow0.3\%$) but notably affects final SRR ($\downarrow5.0\%$). This indicates that verbalizing query intent and constraints provides necessary context for the agent to correctly verify column roles during exploration, preventing false rejections of critical elements.

% \vspace*{-1.5mm}
\paragraph{Dual-Pathway Pruning.} 
In the pruning stage, relying on \textit{Only Selection} or \textit{Only Deletion} reduces SRR by 4.2\% and 16.7\%, respectively. Standard selection misses subtle columns, while aggressive deletion discards necessary structural keys. Combining both pathways, our method achieves the highest SRR of 97.5\%. We note that the average of 383 retained columns corresponds to about 7$k$ tokens, which fits well within the context window for verification.

% \vspace*{-1.5mm}
\paragraph{Agentic Verification.} 
The bottom section confirms that static methods are insufficient. Removing \textit{Agentic Verification} entirely (w/o All) leads to a 21.7\% drop in SRR. Specifically, \textit{Semantic Linking} proves most critical ($\downarrow$14.2\%), as the agent must first hypothesize column roles (e.g., foreign keys) to guide exploration. \textit{Table Exploration} is also vital ($\downarrow$9.2\%) for resolving ambiguities not addressed by metadata. Finally, \textit{Global Synthesis} contributes to robustness ($\downarrow$4.2\%) by ensuring the selected columns form a connected graph.

\begin{table}[t]
\centering
\caption{Ablation study on SQL generation. The evaluation compares performance with and without the guidance module, reporting EX@8 for quality, and average exploration rounds ($\bar{\mathcal{R}}$) and query count ($\bar{\mathcal{Q}}$) per example for efficiency.}
\label{tab:ablation_sql}
\setlength{\tabcolsep}{6.4pt}
\begin{tabular}{llccc}
\toprule
\multirow{2}{*}{\textbf{LLM}} & \multirow{2}{*}{\textbf{Method}} & \multicolumn{3}{c}{\textbf{Spider 2.0-Snow ($N=120$)}} \\
 & & \textbf{EX@8 (\%)$\uparrow$} & $\bar{\mathcal{R}}\downarrow$ & $\bar{\mathcal{Q}}$ \\
\midrule
\multirow{2}{*}{GPT-4o} & w/ Guid. & 29.79 & 3.52 & 10.38 \\
& w/o Guid. & $26.04_{{\color{my_red}\downarrow3.75}}$ & $3.96_{{\color{my_red}\uparrow0.44}}$ & 9.99 \\
\midrule
\multirow{2}{*}{GPT-5} & w/ Guid. & 47.92 & 1.23 & 6.25 \\
& w/o Guid. & $43.65_{\color{my_red}\downarrow4.27}$ & $1.34_{\color{my_red}\uparrow0.11}$ & 6.72 \\
\midrule
\multirow{2}{*}{DeepSeek-R1} & w/ Guid. & 48.75 & 1.32 & 5.04 \\
& w/o Guid. & $47.50_{\color{my_red}\downarrow1.25}$ & $1.69_{\color{my_red}\uparrow0.37}$ & 6.97 \\
\midrule
\multirow{2}{*}{GPT-4.1} & w/ Guid. & 47.19 & 1.62 & 10.57 \\
& w/o Guid. & $46.35_{{\color{my_red}\downarrow0.74}}$ & $1.89_{{\color{my_red}\uparrow0.27}}$ & 10.59 \\
\bottomrule
\end{tabular}
\end{table}

% \vspace{-1.5mm}
\subsubsection{Ablation Study on SQL Generation Components}
\label{analysis:ablation_sql}
While Section \ref{analysis:exploration} demonstrates the impact of exploration, we now further isolate the contribution of the deterministic guidance module by comparing it against a baseline without directives.
We report EX@8 to remove the influence of answer selection. To quantify exploration efficiency, we define the average exploration rounds $\bar{\mathcal{R}}$ and the average query count $\bar{\mathcal{Q}}$ across $m$ questions and $n=8$ samples as:
\begin{equation}
    \bar{\mathcal{R}} = \frac{1}{nm} \sum_{i=1}^m \sum_{j=1}^n \mathcal{R}_{i,j}, \quad \bar{\mathcal{Q}} = \frac{1}{nm} \sum_{i=1}^m \sum_{j=1}^n \mathcal{Q}_{i,j}
\end{equation}
where $\mathcal{R}_{i,j}$ and $\mathcal{Q}_{i,j}$ denote the number of exploration rounds and total exploratory SQL queries generated for the i-$th$ sample in the j-$th$ run. As shown in Table \ref{tab:ablation_sql}, incorporating guidance consistently improves execution accuracy across all models (e.g., GPT-4o $\uparrow3.75\%$). This confirms that guidance helps the agent start with a stronger foundation, leading to more focused and efficient exploration.

Beyond performance gains, the guidance module streamlines the interaction process. Across all models, $\bar{R}$ decreases when guidance is provided, indicating that fewer steps are needed to resolve uncertainties. Notably, while the number of rounds decreases, the total query count $\bar{Q}$ remains comparable or increases, suggesting that the agent performs more meaningful probes per interaction.

\section{Conclusion}

In this work, we present \TheName{}, an agentic framework shifting the Text-to-SQL paradigm from passive schema perception to agentic exploration. For schema linking, we combine logical planning with dual-pathway pruning to navigate massive search spaces, employing agentic exploration to ground column selection in empirical evidence. For SQL generation, we utilize deterministic retrieval to guide data profiling, allowing the agent to resolve implementation details and value ambiguities. Experiments on BIRD and Spider 2.0-Snow show that \TheName{} outperforms competitive baselines. Analysis reveals that agentic exploration acts as a performance multiplier, particularly for stronger models, while deterministic guidance enhances effectiveness and efficiency. We conclude that equipping models to verify hypotheses against physical data is a highly effective mechanism for robust enterprise data analysis.

\bibliographystyle{ACM-Reference-Format}
% \balance
\bibliography{sample-base}


\begin{thebibliography}{33}

%%% the Web address.

\ifx \showCODEN    \undefined \def \showCODEN     #1{\unskip}     \fi
\ifx \showISBNx    \undefined \def \showISBNx     #1{\unskip}     \fi
\ifx \showISBNxiii \undefined \def \showISBNxiii  #1{\unskip}     \fi
\ifx \showISSN     \undefined \def \showISSN      #1{\unskip}     \fi
\ifx \showLCCN     \undefined \def \showLCCN      #1{\unskip}     \fi
\ifx \shownote     \undefined \def \shownote      #1{#1}          \fi
\ifx \showarticletitle \undefined \def \showarticletitle #1{#1}   \fi
\ifx \showURL      \undefined \def \showURL       {\relax}        \fi
% invisible to TeX
\providecommand\bibfield[2]{#2}
\providecommand\bibinfo[2]{#2}
\providecommand\natexlab[1]{#1}
\providecommand\showeprint[2][]{arXiv:#2}

\bibitem[Achiam et~al\mbox{.}(2023)]%
        {GPT4}
\bibfield{author}{\bibinfo{person}{Josh Achiam}, \bibinfo{person}{Steven Adler}, \bibinfo{person}{Sandhini Agarwal}, \bibinfo{person}{Lama Ahmad}, \bibinfo{person}{Ilge Akkaya}, \bibinfo{person}{Florencia~Leoni Aleman}, \bibinfo{person}{Diogo Almeida}, \bibinfo{person}{Janko Altenschmidt}, \bibinfo{person}{Sam Altman}, \bibinfo{person}{Shyamal Anadkat}, {et~al\mbox{.}}} \bibinfo{year}{2023}\natexlab{}.
\newblock \showarticletitle{Gpt-4 technical report}.
\newblock \bibinfo{journal}{\emph{arXiv preprint}} (\bibinfo{year}{2023}).
\newblock
\showeprint{2303.08774}


\bibitem[Agrawal and Nguyen(2025)]%
        {agrawal2025text2sql}
\bibfield{author}{\bibinfo{person}{Sheshansh Agrawal} {and} \bibinfo{person}{Thien Nguyen}.} \bibinfo{year}{2025}\natexlab{}.
\newblock \bibinfo{title}{Open-Sourcing the Best Local Text-to-SQL System}.
\newblock
\urldef\tempurl%
\url{https://contextual.ai/blog/open-sourcing-the-best-local-text-to-sql-system/}
\showURL{%
\tempurl}


\bibitem[Cao et~al\mbox{.}(2024)]%
        {cao2024rsl}
\bibfield{author}{\bibinfo{person}{Zhenbiao Cao}, \bibinfo{person}{Yuanlei Zheng}, \bibinfo{person}{Zhihao Fan}, \bibinfo{person}{Xiaojin Zhang}, \bibinfo{person}{Wei Chen}, {and} \bibinfo{person}{Xiang Bai}.} \bibinfo{year}{2024}\natexlab{}.
\newblock \showarticletitle{Rsl-sql: Robust schema linking in text-to-sql generation}.
\newblock \bibinfo{journal}{\emph{arXiv preprint}} (\bibinfo{year}{2024}).
\newblock
\showeprint{2411.00073}


\bibitem[Deng et~al\mbox{.}(2025)]%
        {deng2025reforce}
\bibfield{author}{\bibinfo{person}{Minghang Deng}, \bibinfo{person}{Ashwin Ramachandran}, \bibinfo{person}{Canwen Xu}, \bibinfo{person}{Lanxiang Hu}, \bibinfo{person}{Zhewei Yao}, \bibinfo{person}{Anupam Datta}, {and} \bibinfo{person}{Hao Zhang}.} \bibinfo{year}{2025}\natexlab{}.
\newblock \showarticletitle{ReFo{RCE}: A Text-to-{SQL} Agent with Self-Refinement, Format Restriction, and Column Exploration}. In \bibinfo{booktitle}{\emph{ICLR 2025 Workshop: VerifAI: AI Verification in the Wild}}.
\newblock
\urldef\tempurl%
\url{https://openreview.net/forum?id=OuFIfDBwQd}
\showURL{%
\tempurl}


\bibitem[Dong et~al\mbox{.}(2023)]%
        {dong2023c3}
\bibfield{author}{\bibinfo{person}{Xuemei Dong}, \bibinfo{person}{Chao Zhang}, \bibinfo{person}{Yuhang Ge}, \bibinfo{person}{Yuren Mao}, \bibinfo{person}{Yunjun Gao}, \bibinfo{person}{Jinshu Lin}, \bibinfo{person}{Dongfang Lou}, {et~al\mbox{.}}} \bibinfo{year}{2023}\natexlab{}.
\newblock \showarticletitle{C3: Zero-shot text-to-sql with chatgpt}.
\newblock \bibinfo{journal}{\emph{arXiv preprint}} (\bibinfo{year}{2023}).
\newblock
\showeprint{2307.07306}


\bibitem[Gao et~al\mbox{.}(2024)]%
        {gao2023text}
\bibfield{author}{\bibinfo{person}{Dawei Gao}, \bibinfo{person}{Haibin Wang}, \bibinfo{person}{Yaliang Li}, \bibinfo{person}{Xiuyu Sun}, \bibinfo{person}{Yichen Qian}, \bibinfo{person}{Bolin Ding}, {and} \bibinfo{person}{Jingren Zhou}.} \bibinfo{year}{2024}\natexlab{}.
\newblock \showarticletitle{Text-to-SQL Empowered by Large Language Models: A Benchmark Evaluation}.
\newblock \bibinfo{journal}{\emph{Proceedings of the VLDB Endowment}} \bibinfo{volume}{17}, \bibinfo{number}{5} (\bibinfo{year}{2024}), \bibinfo{pages}{1132--1145}.
\newblock
\urldef\tempurl%
\url{https://doi.org/10.14778/3641204.3641221}
\showURL{%
\tempurl}


\bibitem[Hao et~al\mbox{.}(2025)]%
        {hao2025text}
\bibfield{author}{\bibinfo{person}{Zhifeng Hao}, \bibinfo{person}{Qibin Song}, \bibinfo{person}{Ruichu Cai}, {and} \bibinfo{person}{Boyan Xu}.} \bibinfo{year}{2025}\natexlab{}.
\newblock \showarticletitle{Text-to-SQL as Dual-State Reasoning: Integrating Adaptive Context and Progressive Generation}.
\newblock \bibinfo{journal}{\emph{arXiv preprint}} (\bibinfo{year}{2025}).
\newblock
\showeprint{2511.21402}


\bibitem[Lee et~al\mbox{.}(2025)]%
        {lee2025mcs}
\bibfield{author}{\bibinfo{person}{Dongjun Lee}, \bibinfo{person}{Choongwon Park}, \bibinfo{person}{Jaehyuk Kim}, {and} \bibinfo{person}{Heesoo Park}.} \bibinfo{year}{2025}\natexlab{}.
\newblock \showarticletitle{Mcs-sql: Leveraging multiple prompts and multiple-choice selection for text-to-sql generation}. In \bibinfo{booktitle}{\emph{Proceedings of the 31st International Conference on Computational Linguistics}}. \bibinfo{pages}{337--353}.
\newblock
\urldef\tempurl%
\url{https://aclanthology.org/2025.coling-main.24/}
\showURL{%
\tempurl}


\bibitem[Lei et~al\mbox{.}(2025)]%
        {lei2024spider}
\bibfield{author}{\bibinfo{person}{Fangyu Lei}, \bibinfo{person}{Jixuan Chen}, \bibinfo{person}{Yuxiao Ye}, \bibinfo{person}{Ruisheng Cao}, \bibinfo{person}{Dongchan Shin}, \bibinfo{person}{Hongjin Su}, \bibinfo{person}{Zhaoqing Suo}, \bibinfo{person}{Hongcheng Gao}, \bibinfo{person}{Wenjing Hu}, \bibinfo{person}{Pengcheng Yin}, {et~al\mbox{.}}} \bibinfo{year}{2025}\natexlab{}.
\newblock \showarticletitle{Spider 2.0: Evaluating language models on real-world enterprise text-to-sql workflows}. In \bibinfo{booktitle}{\emph{International Conference on Learning Representations}}, Vol.~\bibinfo{volume}{2025}. \bibinfo{pages}{28691--28735}.
\newblock
\urldef\tempurl%
\url{https://proceedings.iclr.cc/paper_files/paper/2025/file/46c10f6c8ea5aa6f267bcdabcb123f97-Paper-Conference.pdf}
\showURL{%
\tempurl}


\bibitem[Li et~al\mbox{.}(2024a)]%
        {li2024dawn}
\bibfield{author}{\bibinfo{person}{Boyan Li}, \bibinfo{person}{Yuyu Luo}, \bibinfo{person}{Chengliang Chai}, \bibinfo{person}{Guoliang Li}, {and} \bibinfo{person}{Nan Tang}.} \bibinfo{year}{2024}\natexlab{a}.
\newblock \showarticletitle{The Dawn of Natural Language to SQL: Are We Fully Ready?}
\newblock \bibinfo{journal}{\emph{Proceedings of the VLDB Endowment}} \bibinfo{volume}{17}, \bibinfo{number}{11} (\bibinfo{year}{2024}), \bibinfo{pages}{3318--3331}.
\newblock
\urldef\tempurl%
\url{https://doi.org/10.14778/3681954.3682003}
\showURL{%
\tempurl}


\bibitem[Li et~al\mbox{.}(2025)]%
        {lialpha}
\bibfield{author}{\bibinfo{person}{Boyan Li}, \bibinfo{person}{Jiayi Zhang}, \bibinfo{person}{Ju Fan}, \bibinfo{person}{Yanwei Xu}, \bibinfo{person}{Chong Chen}, \bibinfo{person}{Nan Tang}, {and} \bibinfo{person}{Yuyu Luo}.} \bibinfo{year}{2025}\natexlab{}.
\newblock \showarticletitle{Alpha-{SQL}: Zero-Shot Text-to-{SQL} using Monte Carlo Tree Search}. In \bibinfo{booktitle}{\emph{Forty-second International Conference on Machine Learning}}.
\newblock
\urldef\tempurl%
\url{https://openreview.net/forum?id=kGg1ndttmI}
\showURL{%
\tempurl}


\bibitem[Li et~al\mbox{.}(2023b)]%
        {li2023resdsql}
\bibfield{author}{\bibinfo{person}{Haoyang Li}, \bibinfo{person}{Jing Zhang}, \bibinfo{person}{Cuiping Li}, {and} \bibinfo{person}{Hong Chen}.} \bibinfo{year}{2023}\natexlab{b}.
\newblock \showarticletitle{Resdsql: Decoupling schema linking and skeleton parsing for text-to-sql}. In \bibinfo{booktitle}{\emph{Proceedings of the AAAI Conference on Artificial Intelligence}}, Vol.~\bibinfo{volume}{37}. \bibinfo{pages}{13067--13075}.
\newblock
\urldef\tempurl%
\url{https://doi.org/10.1609/aaai.v37i11.26535}
\showURL{%
\tempurl}


\bibitem[Li et~al\mbox{.}(2024b)]%
        {li2024codes}
\bibfield{author}{\bibinfo{person}{Haoyang Li}, \bibinfo{person}{Jing Zhang}, \bibinfo{person}{Hanbing Liu}, \bibinfo{person}{Ju Fan}, \bibinfo{person}{Xiaokang Zhang}, \bibinfo{person}{Jun Zhu}, \bibinfo{person}{Renjie Wei}, \bibinfo{person}{Hongyan Pan}, \bibinfo{person}{Cuiping Li}, {and} \bibinfo{person}{Hong Chen}.} \bibinfo{year}{2024}\natexlab{b}.
\newblock \showarticletitle{Codes: Towards building open-source language models for text-to-sql}.
\newblock \bibinfo{journal}{\emph{Proceedings of the ACM on Management of Data}} \bibinfo{volume}{2}, \bibinfo{number}{3} (\bibinfo{year}{2024}), \bibinfo{pages}{1--28}.
\newblock
\urldef\tempurl%
\url{https://doi.org/10.1145/3654930}
\showURL{%
\tempurl}


\bibitem[Li et~al\mbox{.}(2023a)]%
        {li2023can}
\bibfield{author}{\bibinfo{person}{Jinyang Li}, \bibinfo{person}{Binyuan Hui}, \bibinfo{person}{Ge Qu}, \bibinfo{person}{Jiaxi Yang}, \bibinfo{person}{Binhua Li}, \bibinfo{person}{Bowen Li}, \bibinfo{person}{Bailin Wang}, \bibinfo{person}{Bowen Qin}, \bibinfo{person}{Ruiying Geng}, \bibinfo{person}{Nan Huo}, {et~al\mbox{.}}} \bibinfo{year}{2023}\natexlab{a}.
\newblock \showarticletitle{Can llm already serve as a database interface? a big bench for large-scale database grounded text-to-sqls}.
\newblock \bibinfo{journal}{\emph{Advances in Neural Information Processing Systems}}  \bibinfo{volume}{36} (\bibinfo{year}{2023}), \bibinfo{pages}{42330--42357}.
\newblock
\urldef\tempurl%
\url{https://proceedings.neurips.cc/paper_files/paper/2023/file/83fc8fab1710363050bbd1d4b8cc0021-Paper-Datasets_and_Benchmarks.pdf}
\showURL{%
\tempurl}


\bibitem[Liao et~al\mbox{.}(2025)]%
        {liao2025learnat}
\bibfield{author}{\bibinfo{person}{Weibin Liao}, \bibinfo{person}{Xin Gao}, \bibinfo{person}{Tianyu Jia}, \bibinfo{person}{Rihong Qiu}, \bibinfo{person}{Yifan Zhu}, \bibinfo{person}{Yang Lin}, \bibinfo{person}{Xu Chu}, \bibinfo{person}{Junfeng Zhao}, {and} \bibinfo{person}{Yasha Wang}.} \bibinfo{year}{2025}\natexlab{}.
\newblock \showarticletitle{LearNAT: Learning NL2SQL with AST-guided Task Decomposition for Large Language Models}.
\newblock \bibinfo{journal}{\emph{arXiv preprint}} (\bibinfo{year}{2025}).
\newblock
\showeprint{2504.02327}


\bibitem[Liu et~al\mbox{.}(2024)]%
        {liu2024deepseek}
\bibfield{author}{\bibinfo{person}{Aixin Liu}, \bibinfo{person}{Bei Feng}, \bibinfo{person}{Bing Xue}, \bibinfo{person}{Bingxuan Wang}, \bibinfo{person}{Bochao Wu}, \bibinfo{person}{Chengda Lu}, \bibinfo{person}{Chenggang Zhao}, \bibinfo{person}{Chengqi Deng}, \bibinfo{person}{Chenyu Zhang}, \bibinfo{person}{Chong Ruan}, {et~al\mbox{.}}} \bibinfo{year}{2024}\natexlab{}.
\newblock \showarticletitle{Deepseek-v3 technical report}.
\newblock \bibinfo{journal}{\emph{arXiv preprint}} (\bibinfo{year}{2024}).
\newblock
\showeprint{2412.19437}


\bibitem[Liu et~al\mbox{.}(2025)]%
        {liu2025survey}
\bibfield{author}{\bibinfo{person}{Xinyu Liu}, \bibinfo{person}{Shuyu Shen}, \bibinfo{person}{Boyan Li}, \bibinfo{person}{Peixian Ma}, \bibinfo{person}{Runzhi Jiang}, \bibinfo{person}{Yuxin Zhang}, \bibinfo{person}{Ju Fan}, \bibinfo{person}{Guoliang Li}, \bibinfo{person}{Nan Tang}, {and} \bibinfo{person}{Yuyu Luo}.} \bibinfo{year}{2025}\natexlab{}.
\newblock \showarticletitle{A Survey of Text-to-SQL in the Era of LLMs: Where are we, and where are we going?}
\newblock \bibinfo{journal}{\emph{IEEE Transactions on Knowledge and Data Engineering}} (\bibinfo{year}{2025}).
\newblock
\urldef\tempurl%
\url{https://doi.ieeecomputersociety.org/10.1109/TKDE.2025.3592032}
\showURL{%
\tempurl}


\bibitem[Pourreza et~al\mbox{.}(2025)]%
        {pourreza2024chase}
\bibfield{author}{\bibinfo{person}{Mohammadreza Pourreza}, \bibinfo{person}{Hailong Li}, \bibinfo{person}{Ruoxi Sun}, \bibinfo{person}{Yeounoh Chung}, \bibinfo{person}{Shayan Talaei}, \bibinfo{person}{Gaurav~Tarlok Kakkar}, \bibinfo{person}{Yu Gan}, \bibinfo{person}{Amin Saberi}, \bibinfo{person}{Fatma Ozcan}, {and} \bibinfo{person}{Sercan Arik}.} \bibinfo{year}{2025}\natexlab{}.
\newblock \showarticletitle{Chase-sql: Multi-path reasoning and preference optimized candidate selection in text-to-sql}. In \bibinfo{booktitle}{\emph{International Conference on Learning Representations}}, Vol.~\bibinfo{volume}{2025}. \bibinfo{pages}{60385--60415}.
\newblock
\urldef\tempurl%
\url{https://proceedings.iclr.cc/paper_files/paper/2025/file/974ff7b5bf08dbf9400b5d599a39c77f-Paper-Conference.pdf}
\showURL{%
\tempurl}


\bibitem[Pourreza and Rafiei(2023)]%
        {pourreza2023din}
\bibfield{author}{\bibinfo{person}{Mohammadreza Pourreza} {and} \bibinfo{person}{Davood Rafiei}.} \bibinfo{year}{2023}\natexlab{}.
\newblock \showarticletitle{Din-sql: Decomposed in-context learning of text-to-sql with self-correction}.
\newblock \bibinfo{journal}{\emph{Advances in Neural Information Processing Systems}}  \bibinfo{volume}{36} (\bibinfo{year}{2023}), \bibinfo{pages}{36339--36348}.
\newblock
\urldef\tempurl%
\url{https://proceedings.neurips.cc/paper_files/paper/2023/file/72223cc66f63ca1aa59edaec1b3670e6-Paper-Conference.pdf}
\showURL{%
\tempurl}


\bibitem[Qu et~al\mbox{.}(2024)]%
        {qu2024before}
\bibfield{author}{\bibinfo{person}{Ge Qu}, \bibinfo{person}{Jinyang Li}, \bibinfo{person}{Bowen Li}, \bibinfo{person}{Bowen Qin}, \bibinfo{person}{Nan Huo}, \bibinfo{person}{Chenhao Ma}, {and} \bibinfo{person}{Reynold Cheng}.} \bibinfo{year}{2024}\natexlab{}.
\newblock \showarticletitle{Before Generation, Align it! A Novel and Effective Strategy for Mitigating Hallucinations in Text-to-SQL Generation}. In \bibinfo{booktitle}{\emph{Findings of the Association for Computational Linguistics ACL 2024}}. \bibinfo{pages}{5456--5471}.
\newblock
\urldef\tempurl%
\url{https://aclanthology.org/2024.findings-acl.324/}
\showURL{%
\tempurl}


\bibitem[Shi et~al\mbox{.}(2025)]%
        {shi2025survey}
\bibfield{author}{\bibinfo{person}{Liang Shi}, \bibinfo{person}{Zhengju Tang}, \bibinfo{person}{Nan Zhang}, \bibinfo{person}{Xiaotong Zhang}, {and} \bibinfo{person}{Zhi Yang}.} \bibinfo{year}{2025}\natexlab{}.
\newblock \showarticletitle{A survey on employing large language models for text-to-sql tasks}.
\newblock \bibinfo{journal}{\emph{Comput. Surveys}} \bibinfo{volume}{58}, \bibinfo{number}{2} (\bibinfo{year}{2025}), \bibinfo{pages}{1--37}.
\newblock
\urldef\tempurl%
\url{https://doi.org/10.1145/3737873}
\showURL{%
\tempurl}


\bibitem[Shkapenyuk et~al\mbox{.}(2025)]%
        {shkapenyuk2025automatic}
\bibfield{author}{\bibinfo{person}{Vladislav Shkapenyuk}, \bibinfo{person}{Divesh Srivastava}, \bibinfo{person}{Theodore Johnson}, {and} \bibinfo{person}{Parisa Ghane}.} \bibinfo{year}{2025}\natexlab{}.
\newblock \showarticletitle{Automatic Metadata Extraction for Text-to-SQL}.
\newblock \bibinfo{journal}{\emph{arXiv preprint}} (\bibinfo{year}{2025}).
\newblock
\showeprint{2505.19988}


\bibitem[Sun and Chen(2026)]%
        {sun2026cacherag}
\bibfield{author}{\bibinfo{person}{Yushi Sun} {and} \bibinfo{person}{Lei Chen}.} \bibinfo{year}{2026}\natexlab{}.
\newblock \showarticletitle{CacheRAG: A Semantic Caching System for Retrieval-Augmented Generation in Knowledge Graph Question Answering}.
\newblock \bibinfo{journal}{\emph{arXiv preprint}} (\bibinfo{year}{2026}).
\newblock
\showeprint{2604.26176}


\bibitem[Sun et~al\mbox{.}(2026)]%
        {sun2026lakehopper}
\bibfield{author}{\bibinfo{person}{Yushi Sun}, \bibinfo{person}{Xujia Li}, \bibinfo{person}{Nan Tang}, \bibinfo{person}{Quanqing Xu}, \bibinfo{person}{Chuanhui Yang}, {and} \bibinfo{person}{Lei Chen}.} \bibinfo{year}{2026}\natexlab{}.
\newblock \showarticletitle{LakeHopper: Cross Data Lakes Column Type Annotation through Model Adaptation}.
\newblock \bibinfo{journal}{\emph{arXiv preprint}} (\bibinfo{year}{2026}).
\newblock
\showeprint{2602.08793}


\bibitem[Sun et~al\mbox{.}(2025)]%
        {sun2025kerag}
\bibfield{author}{\bibinfo{person}{Yushi Sun}, \bibinfo{person}{Kai Sun}, \bibinfo{person}{Yifan~Ethan Xu}, \bibinfo{person}{Xiao Yang}, \bibinfo{person}{Xin~Luna Dong}, \bibinfo{person}{Nan Tang}, {and} \bibinfo{person}{Lei Chen}.} \bibinfo{year}{2025}\natexlab{}.
\newblock \showarticletitle{KERAG: Knowledge-Enhanced Retrieval-Augmented Generation for Advanced Question Answering}. In \bibinfo{booktitle}{\emph{Findings of the Association for Computational Linguistics: EMNLP 2025}}. \bibinfo{pages}{6194--6216}.
\newblock
\urldef\tempurl%
\url{https://aclanthology.org/2025.findings-emnlp.329/}
\showURL{%
\tempurl}


\bibitem[Sun et~al\mbox{.}(2023)]%
        {sun2023reca}
\bibfield{author}{\bibinfo{person}{Yushi Sun}, \bibinfo{person}{Hao Xin}, {and} \bibinfo{person}{Lei Chen}.} \bibinfo{year}{2023}\natexlab{}.
\newblock \showarticletitle{RECA: Related Tables Enhanced Column Semantic Type Annotation Framework}.
\newblock \bibinfo{journal}{\emph{Proceedings of the VLDB Endowment}} \bibinfo{volume}{16}, \bibinfo{number}{6} (\bibinfo{year}{2023}), \bibinfo{pages}{1319--1331}.
\newblock
\urldef\tempurl%
\url{https://doi.org/10.14778/3583140.3583149}
\showURL{%
\tempurl}


\bibitem[Talaei et~al\mbox{.}(2024)]%
        {talaei2024chess}
\bibfield{author}{\bibinfo{person}{Shayan Talaei}, \bibinfo{person}{Mohammadreza Pourreza}, \bibinfo{person}{Yu-Chen Chang}, \bibinfo{person}{Azalia Mirhoseini}, {and} \bibinfo{person}{Amin Saberi}.} \bibinfo{year}{2024}\natexlab{}.
\newblock \showarticletitle{Chess: Contextual harnessing for efficient sql synthesis}.
\newblock \bibinfo{journal}{\emph{arXiv preprint}} (\bibinfo{year}{2024}).
\newblock
\showeprint{2405.16755}


\bibitem[Wang et~al\mbox{.}(2025)]%
        {wang2025mac}
\bibfield{author}{\bibinfo{person}{Bing Wang}, \bibinfo{person}{Changyu Ren}, \bibinfo{person}{Jian Yang}, \bibinfo{person}{Xinnian Liang}, \bibinfo{person}{Jiaqi Bai}, \bibinfo{person}{Linzheng Chai}, \bibinfo{person}{Zhao Yan}, \bibinfo{person}{Qian-Wen Zhang}, \bibinfo{person}{Di Yin}, \bibinfo{person}{Xing Sun}, {et~al\mbox{.}}} \bibinfo{year}{2025}\natexlab{}.
\newblock \showarticletitle{Mac-sql: A multi-agent collaborative framework for text-to-sql}. In \bibinfo{booktitle}{\emph{Proceedings of the 31st International Conference on Computational Linguistics}}. \bibinfo{pages}{540--557}.
\newblock
\urldef\tempurl%
\url{https://aclanthology.org/2025.coling-main.36/}
\showURL{%
\tempurl}


\bibitem[Wang et~al\mbox{.}(2026)]%
        {wang2025autolink}
\bibfield{author}{\bibinfo{person}{Ziyang Wang}, \bibinfo{person}{Yuanlei Zheng}, \bibinfo{person}{Zhenbiao Cao}, \bibinfo{person}{Xiaojin Zhang}, \bibinfo{person}{Zhongyu Wei}, \bibinfo{person}{Pei Fu}, \bibinfo{person}{Zhenbo Luo}, \bibinfo{person}{Wei Chen}, {and} \bibinfo{person}{Xiang Bai}.} \bibinfo{year}{2026}\natexlab{}.
\newblock \showarticletitle{AutoLink: Autonomous Schema Exploration and Expansion for Scalable Schema Linking in Text-to-SQL at Scale}. In \bibinfo{booktitle}{\emph{Proceedings of the AAAI Conference on Artificial Intelligence}}, Vol.~\bibinfo{volume}{40}. \bibinfo{pages}{33809--33817}.
\newblock
\urldef\tempurl%
\url{https://doi.org/10.1609/aaai.v40i40.40672}
\showURL{%
\tempurl}


\bibitem[Xie et~al\mbox{.}(2024)]%
        {xie2024mag}
\bibfield{author}{\bibinfo{person}{Wenxuan Xie}, \bibinfo{person}{Gaochen Wu}, {and} \bibinfo{person}{Bowen Zhou}.} \bibinfo{year}{2024}\natexlab{}.
\newblock \showarticletitle{Mag-sql: Multi-agent generative approach with soft schema linking and iterative sub-sql refinement for text-to-sql}.
\newblock \bibinfo{journal}{\emph{arXiv preprint}} (\bibinfo{year}{2024}).
\newblock
\showeprint{2408.07930}


\bibitem[Xie et~al\mbox{.}(2025)]%
        {xie2025opensearchsqlenhancingtexttosqldynamic}
\bibfield{author}{\bibinfo{person}{Xiangjin Xie}, \bibinfo{person}{Guangwei Xu}, \bibinfo{person}{Lingyan Zhao}, {and} \bibinfo{person}{Ruijie Guo}.} \bibinfo{year}{2025}\natexlab{}.
\newblock \showarticletitle{Opensearch-sql: Enhancing text-to-sql with dynamic few-shot and consistency alignment}.
\newblock \bibinfo{journal}{\emph{Proceedings of the ACM on Management of Data}} \bibinfo{volume}{3}, \bibinfo{number}{3} (\bibinfo{year}{2025}), \bibinfo{pages}{1--24}.
\newblock
\urldef\tempurl%
\url{https://doi.org/10.1145/3725331}
\showURL{%
\tempurl}


\bibitem[Yang et~al\mbox{.}(2025)]%
        {yang2025qwen3}
\bibfield{author}{\bibinfo{person}{An Yang}, \bibinfo{person}{Anfeng Li}, \bibinfo{person}{Baosong Yang}, \bibinfo{person}{Beichen Zhang}, \bibinfo{person}{Binyuan Hui}, \bibinfo{person}{Bo Zheng}, \bibinfo{person}{Bowen Yu}, \bibinfo{person}{Chang Gao}, \bibinfo{person}{Chengen Huang}, \bibinfo{person}{Chenxu Lv}, {et~al\mbox{.}}} \bibinfo{year}{2025}\natexlab{}.
\newblock \showarticletitle{Qwen3 technical report}.
\newblock \bibinfo{journal}{\emph{arXiv preprint}} (\bibinfo{year}{2025}).
\newblock
\showeprint{2505.09388}


\bibitem[Yu et~al\mbox{.}(2018)]%
        {yu2018spider}
\bibfield{author}{\bibinfo{person}{Tao Yu}, \bibinfo{person}{Rui Zhang}, \bibinfo{person}{Kai Yang}, \bibinfo{person}{Michihiro Yasunaga}, \bibinfo{person}{Dongxu Wang}, \bibinfo{person}{Zifan Li}, \bibinfo{person}{James Ma}, \bibinfo{person}{Irene Li}, \bibinfo{person}{Qingning Yao}, \bibinfo{person}{Shanelle Roman}, {et~al\mbox{.}}} \bibinfo{year}{2018}\natexlab{}.
\newblock \showarticletitle{Spider: A large-scale human-labeled dataset for complex and cross-domain semantic parsing and text-to-sql task}. In \bibinfo{booktitle}{\emph{Proceedings of the 2018 conference on empirical methods in natural language processing}}. \bibinfo{pages}{3911--3921}.
\newblock
\urldef\tempurl%
\url{https://aclanthology.org/D18-1425/}
\showURL{%
\tempurl}


\end{thebibliography}

\appendix
\section{Instruction Templates}
\label{app:prompts}

This section provides the instruction templates referenced in the main text. Each prompt corresponds to specific components of our framework, as detailed below:

\begin{itemize}
[leftmargin=*,itemsep=0pt,parsep=0.2em,topsep=0.3em,partopsep=0.3em]
    \item $F^{sl}_{plan}$ and $F^{sl}_{agg}$: For Logical Planning in Section \ref{sec: logical plan}.
    
    \item $F^{sl}_{del}$ and $F^{sl}_{sel}$: For Dual-Pathway Pruning in Section \ref{sec:pruning}.
    
    \item $F^{sl}_{semantics}$, $F^{sl}_{exp}$, and $F^{sl}_{final}$: For Agentic Exploration in Section \ref{sec:agentic_exploration}.
    
    \item $F^{sql}_{kw}$: For SQL realization path generation in Section \ref{sec:derterminitic_retrieval}.

    \item $\mathcal{A}$: Defining Action Space in Section \ref{sec:sql_exploration}.
    \item The instructions for Evidence linking and model-based Answer Selection (as a tie-breaker) are discussed in Section~\ref{sec:sql_details}.
\end{itemize}

\begin{promptbox}[Logical Planning ($F^{sl}_{plan}$)]
*** TASK CONTEXT ***

You are a Lead Data Architect. Your task is to break down the User Question into abstract logical steps needed to answer it.
\\

**IMPORTANT**: Do NOT reference specific table or column names yet. Focus purely on the logic (e.g., filter, join, count, aggregate).
\\

*** USER QUESTION ***

\{question\}
\\

*** OUTPUT FORMAT ***

\{

\quad"logical\_steps": [
    
\quad\quad"1. Identify [Entity]...",
        
\quad\quad"2. Filter where [Condition]...",
        
\quad\quad"3. Link [Entity A] to [Entity B]...",
        
\quad\quad"4. Calculate [Aggregation]..."
        
\quad]
    
\}
\end{promptbox}

\begin{promptbox}[Aggregating Plan Candidates ($F^{sl}_{agg}$)]
*** TASK CONTEXT ***

We have collected \{len(candidates)\} draft logical plans. Synthesize them into a single, comprehensive Master Logical Plan.
Ensure the steps cover all conditions, filters, joins, and aggregations required.
\\

*** USER QUESTION ***

\{question\}
\\

*** DRAFT PLANS ***

\{plans\_text\}
\\

Output just the steps as a numbered list.
\end{promptbox}

\begin{promptbox}[Identifying Deletion Set ($F^{sl}_{del}$)]
*** TASK CONTEXT ***

You are a Lead Data Architect. You have a Logical Plan to answer a query.

Your task: **Negative Pruning**. Identify database tables or columns that are **100\% IRRELEVANT** to the plan.
\\

*** USER QUESTION ***

\{question\}
\\

*** MASTER LOGICAL PLAN ***

\{logical\_plan\}
\\

*** FULL DATABASE SCHEMA ***

\{schema\}
\\

*** EVIDENCE ***

\{evidence\}
\\
            
*** STRICT GUIDELINES ***

1. **High Recall (Safety)**:

\quad- If the column name is related to the query (even 1\% chance), you should keep it. If not, check the desciption to see if it is related to the query. Sometimes the description is not clear, then you should pay close attention to the sample rows of the table. If the sample values of some columns are related to the query, you should keep these columns. If all of these information are not clear enough, remove it.

2. **Definition of Relevance**: Relevance includes both **Lexical Matching** and **Semantic Relatedness** over column name and description.

\quad- **Lexical**: If a word from the query appears in the name (e.g., query mentions "school" -> keep `school\_code`, `school\_type`, etc.), it MUST be retained.

\quad- **Semantic**: Keep columns conceptually related to the topic. For example, if the query asks about "patents that were granted in ...", then the column `grant\_date` should be kept.

\quad- **CRITICAL**: Do NOT remove discriminator columns such as `xxx\_id`, `xxx\_name`, `xxx\_code`, or `xxx\_type` if the table itself is kept.

3. **Output Removal List**:

\quad- **Tables**: If a whole table is irrelevant, list it in `obviously\_irrelevant\_tables`. Then all columns of that table will be kept. You do NOT need to list its columns separately.

\quad- **Columns**: If specific columns of a table are noise, list them in `obviously\_irrelevant\_columns`.

4. **Grouped Tables**: If multiple tables are presented as sharing the same columns, you MUST list the removal instructions for **EACH** table explicitly. Pay close attention to name differences within the group (e.g., xx\_2017 vs xx\_2026), as these reflect specific data dimensions (like time) that determine relevance to the query.
\\

*** OUTPUT FORMAT ***

\verb|```|json

\{

\quad"obviously\_irrelevant\_tables": ["table\_unused\_1", "table\_unused\_2"],

\quad"obviously\_irrelevant\_columns": [

\quad\quad\{

\quad\quad\quad"table": "t1", 

\quad\quad\quad"columns": ["col\_unused\_1", "col\_unused\_2"]

\quad\quad\}

\quad]

\}

\verb|```|
\end{promptbox}

\begin{promptbox}[Identifying Preservation Set ($F^{sl}_{sel}$)]
*** TASK CONTEXT ***

You are a Lead Data Architect. You have a Logical Plan to answer a query.

Your task: **Positive Selection**. Identify database tables or columns that are **RELEVANT** or **NECESSARY** to the plan.
\\

*** USER QUESTION ***

{question}
\\

*** MASTER LOGICAL PLAN 

\{logical\_plan\}
\\

*** FULL DATABASE SCHEMA ***

\{schema\}
\\        

*** EVIDENCE ***

\{evidence\}
\\
            
*** STRICT GUIDELINES ***

1. **High Recall (Safety)**: Select ALL columns that might be useful for joining, filtering, grouping, or returning results. If you are not sure about the relevance of a column, e.g., the name and the description are ambiguous, **PICK IT**.

2. **Definition of Relevance**: Relevance includes both **Lexical Matching** and **Semantic Relatedness** over column name and description.

\quad- **Lexical**: If a word from the query appears in the table or column name (e.g., query mentions "school" -> keep `school\_code`, `school\_type`, etc.), it MUST be selected.

\quad- **Semantic**: Identify columns conceptually related to the topic. For example, if the query asks about "patents that were granted in ...", then the column `grant\_date` should be kept.

\quad- **Discriminators**: ALWAYS select primary keys and common identifiers (`xxx\_id`, `xxx\_code`, `xxx\_name`) for relevant tables, as they are needed for joins.

3. **Output Selection List**:

\quad- **Tables**: If a whole table is relevant, list it in `relevant\_tables`.

\quad- **Columns**: List specific useful columns in `relevant\_columns`. If a table is already listed in `relevant\_tables`, the columns can be omitted.

4. **Grouped Tables**: If multiple tables are presented as sharing the same columns, you MUST list the selection instructions for **EACH** table explicitly. Pay close attention to name differences within the group (e.g., xx\_2017 vs xx\_2026), as these reflect specific data dimensions (like time) that determine relevance to the query.
\\

*** OUTPUT FORMAT ***

\verb|```|json

\{

\quad"relevant\_tables": ["table\_useful\_1"],

\quad"relevant\_columns": [

\quad\quad\{

\quad\quad\quad"table": "t1", 

\quad\quad\quad"columns": ["col\_useful\_1", "col\_pk\_id"]

\quad\quad\}

\quad]

\}

\verb|```|
\end{promptbox}

\begin{promptbox}[Semantic Linking ($F^{sl}_{semantics}$)]
*** TASK CONTEXT ***

You are a Senior Data Architect. You have full visibility of the database schema and a user question.

Your goal is to perform **Semantic Linking**: Analyze the database structure and how it grounds the user's intent.

\{CRITICAL\_RULES\}
\\

*** USER QUESTION ***

\{question\}
\\

*** Logical Plan ***

\{logical\_plan\}
\\

*** Evidence ***

\{evidence\}
\\

*** DATABASE SCHEMA ***

\{schema\}
\\

*** YOUR TASKS ***

1. **Database Structure Overview**: Describe the database structure in detail (e.g., 'A banking system with customers and transactions...').

2. **Query-Specific Content Analysis**: Analyze the query against the available columns. Identify which columns are likely targets, filters, or join keys.

3. **Table Functional Analysis**: For EVERY potentially relevant table, describe its specific function regarding this query.

\quad- Is it a **Target Table**? (Contains the answer columns)
   
\quad- Is it a **Bridge Table**? (Doesn't have semantic data but is needed to join Table A and Table B via Foreign Keys)
   
\quad- Is it a **Filtering Table**? (Contains columns for WHERE clauses)
   
\quad- **CRITICAL**: A table may have multiple roles. If a table is needed as a BRIDGE, you MUST explicitly state that it connects Entity X and Entity Y, even if it looks empty of content.
\\

*** OUTPUT FORMAT ***

\{

\quad"database\_structure": "Database structure overview...",

\quad"query\_specific\_content\_analysis": "Detailed mapping of query terms to DB columns/logic...",

\quad"table\_functions": \{

\quad\quad"table\_name\_1": "Acts as a bridge table connecting Students and Classes via student\_id and class\_id.",

\quad\quad"table\_name\_2": "Contains the 'score' column needed for calculation and 'exam\_date' for filtering."

\quad\}

\}

Perform the semantic linking analysis:
\end{promptbox}

\begin{promptbox}[Parallel Data Profiling ($F^{sl}_{exp}$)]
*** TASK CONTEXT ***

You are an agent exploring a database table to verify its relevance to a user question.

You must not explore randomly. You must verify if this table fits its anticipated role.
            
\{CRITICAL\_RULES\}
\\

*** TARGET TABLE: \{table\_name\} ***

Columns:

\{column\_name\} (\{column\_type\}): \{column\_description\}

...
\\

*** USER QUESTION ***

\{question\}
\\

*** ANTICIPATED ROLE ***

This table was identified as: \{semantic\_role\}. Use this to guide your exploration.
\\

*** Evidence ***

\{evidence\}
\\

*** YOUR MISSION ***

Generate 3-8 SQLite queries to investigate. **Focus on understanding the table's semantics and utility.**
\\

**Motivation for Exploration**:

1. **Semantic Alignment**: Check distinct values to understand what the column *means* versus what the query *needs*. (e.g., If column is 'type', does it contain the specific categories? If 'status', does it contain values like 'Active' or code '1'?)

2. **Granularity \& Scope**: Verify the table's grain (e.g., is it one row per Order or per Item?). This determines if it supports the required aggregations.

3. **Bridge/Connectivity**: If this looks like a linking table, verify the Foreign Keys are populated (not all NULL) to ensure it can actually serve as a bridge.

4. **Data Quality**: Are critical columns (targets for filters or answers) usable, or are they mostly NULL?
\\

*** OUTPUT FORMAT ***

Provide SQL queries in a single `sql` block with comments explaining the *motivation*.

\verb|```|sql

-- Motivation: Checking distinct values in 'status' to see if it aligns with the query's filter requirement

SELECT DISTINCT status FROM table\_name LIMIT 10;

\verb|```|
\\

Generate your exploration queries:
\\

[\textit{After Profiling ...}]
\\

*** TASK ***

Based on the exploration history and results above, determine if table `\{table\_name\}` is RELEVANT to the User Question.
\\

*** EXPLORATION EVIDENCE ***

\{observations\}
\\

*** USER QUESTION ***

\{question\}
\\

*** EVIDENCE ***

\{evidence\}
\\

*** DECISION GUIDELINES ***

- **Direct Match**: Contains the specific answer data.

- **Bridge Table**: Contains IDs needed to join other relevant tables (CRITICAL: Keep even if no other useful data).

- **Filter Source**: Contains columns needed to restrict the result.

- **Calculation Support**: Contains numerical columns needed for aggregation (e.g., `score` for AVG, `price` for SUM).
\\

*** OUTPUT GUIDELINES ***

- `relevance\_reason`: Explain the LOGICAL role (e.g., 'Provides the Join Key for X and Y', 'Contains the target column Z').

- `observations`: Summarize FACTUAL findings from exploration (e.g., 'Column A contains integer codes 1-5', 'Table is empty').

- `table\_summary`: A concise summary of what this table represents in the context of the query.
\\

*** OUTPUT FORMAT ***

\verb|```|json

\{

\quad"relevant": true/false,

\quad"relevant\_columns": [

\quad\{

\quad\quad"column\_name": "name",

\quad\quad"relevance\_reason": "...",

\quad\quad"observations": "..."

\quad\}

\quad],

\quad"table\_summary": "..."

\}

\verb|```|

Provide your analysis:
\end{promptbox}

\begin{promptbox}[Global Synthesis ($F^{sl}_{final}$)]
*** TASK CONTEXT ***

You are the Lead Data Architect. We are synthesizing initial exploration findings.

Review the [MARKED RELEVANT] and [MARKED IRRELEVANT] tables. Fix blind spots.
\\

*** USER QUESTION ***

\{question\}
\\

*** EVIDENCE ***

\{question\}
\\

*** SEMANTIC ANALYSIS ***

\{db\_summary\}
\\

*** SCHEMA STATUS ***

Table: \{t\_name\} [MARKED RELEVANT/MARKED IRRELEVANT]

Columns:

\{column\_name\} (\{column\_type\}): \{column\_description\}

Observations: \{observations\}

Reason: \{reason\}
\\

*** YOUR MISSION ***

Determine the final list of columns required to write the SQL query.

You must ensure the selected columns form a connected graph (tables can be joined) and cover all functional requirements of the query.
\\
        
*** SELECTION CRITERIA (FUNCTIONALITY) ***

Keep a column if it serves one of the following purposes:

1. **Identification**: Unique identifiers (IDs, Codes) needed to count or distinguish entities (Primary keys).

2. **Linking**: Columns needed to join two tables together (Foreign Keys).

3. **Filtering**: Columns involved in conditions (e.g., status='Active', date > 2023).

4. **Aggregation**: Numerical columns for calculations (Sum, Avg, Max, Min).

5. **Grouping \& Sorting**: Columns used for 'GROUP BY' or 'ORDER BY'.

6. **Direct Result**: Columns explicitly requested in the output.

**Note on Multi-Path**: If multiple columns might serve the same purpose, KEEP ALL OF THEM. Alternative columns might help to construct another solution paths.

**Note on Type of Entity**: DO NOT guess the type of an unspecified entity even you have some prior knowledge, e.g., if the query contains location entity like 'Riverside', then ALL columns related to location (e.g., County, District, etc.) should be kept. Another example is 'Fresno County Office of Education' which is actually a full name of a district.
\\

*** REJECTION REQUIREMENTS ***

If a column was marked as **[MARKED RELEVANT]** in the Schema Status but you decide to **REJECT** it, you MUST include it in the `rejected\_candidates` list with a `reject\_reason` explaining why it is unnecessary. You can NOT reject a column for the reason that it is only a potentially useful column.
\\

*** INTERACTIVE PROCESS ***

You can perform up to {MAX\_REFINE\_ROUNDS} rounds of verification.

- To EXPLORE: Output `exploration\_queries` in JSON to test joins or content.

- To FINISH: Output `[CONFIRM]` in the JSON (or just output the final refined\_schema without queries).
\\

*** OUTPUT FORMAT ***

You MUST explicitly list rejected candidates to prove you considered them.

**IMPORTANT**: In 'rejected\_candidates', ONLY list columns that were previously marked RELEVANT but you decided to reject, OR columns that look ambiguous. Do NOT list obviously irrelevant columns to save space.
\\

\verb|```|json

\{

\quad"refined\_schema": \{

\quad\quad"table\_name": \{ 

\quad\quad\quad"relevant\_columns": [

\quad\quad\quad\quad\{

\quad\quad\quad\quad\quad"column\_name": "...", 

\quad\quad\quad\quad\quad"relevance\_reason": "Functional reason (e.g., Needed for Filtering)"

\quad\quad\quad\quad\}

\quad\quad\quad]

\quad\quad\}

\quad\},

\quad"rejected\_candidates": [

\quad\quad\{

\quad\quad\quad"table": "t1",

\quad\quad\quad"column": "c1", 

\quad\quad\quad"reject\_reason": "Originally marked relevant, but rejected because..."

\quad\quad\}

\quad],

\quad"exploration\_queries": ["SELECT 1 FROM t1 JOIN t2 ON t1.id=t2.id LIMIT 1"],

\quad"status": "EXPLORING" or "[CONFIRM]"

\}

\verb|```|
\\

Begin refinement:
\end{promptbox}

\begin{promptbox}[Inferring potential SQL realization paths ($F^{sql}_{kw}$)]
You are refining a logical plan. For each step, think about:
1. What information is needed

2. Different ways to obtain it (direct access, join, calculation, etc.)

3. Keywords that describe the operation
\\

QUESTION: \{question\}
        
Evidence: \{Evidence\}

Schema: \{Schema\}

CURRENT PLAN: \{Logical Plan\}
\\

YOUR TASK:

Refine the plan by analyzing each step. For each step, provide:
        
Step N: [Brief description]

  - Info need: [What information is required]
  
  - Possible paths: [List 2-3 ways to get this info, e.g., 'direct column X', 'join tables A-B', 'calculate using formula']
  
  - Keywords: [table names, column names, operations like filter/join/aggregate, concepts]
  
  - Evidence: [exact evidence text if applicable]
\\

EXAMPLE:

Step 1: Filter for high schools

  - Info need: Identify high school records
  
  - Possible paths: 'school\_type column', 'EILCode column', 'join with school\_types table'
  
  - Keywords: schools, school\_type, EILCode, filter, high school
  
  - Evidence: EILCode = 'HS' means high school
\\

Step 2: Calculate average score

  - Info need: Average of scores
  
  - Possible paths: 'AVG(score\_column)', 'SUM/COUNT formula', 'pre-computed avg\_score column'
  
  - Keywords: scores, average, AVG, aggregate, calculation
\\

IMPORTANT:

- Focus ONLY on the logical steps needed to answer the question

- Do NOT specify output columns in this plan

- Evidence: preserve EXACTLY (formulas, column names, values)

- Paths: list alternatives naturally (don't force if only one way makes sense)

- Keywords: comprehensive but relevant

- Keep plan abstract (avoid specific table/column names unless from evidence)
\\

Now refine the plan:
\end{promptbox}

\begin{promptbox}[Action Space Definition ($\mathcal{A}$)]
You are an expert SQL query generator. Your task is to convert natural language questions into SQL queries.
\\

\# AVAILABLE ACTIONS

**CRITICAL**: Always start your response with EXACTLY ONE action tag ([EXPLORE], [REFINE], [SQL], or [CONFIRM]) at the very beginning.
\\
            
\#\# [EXPLORE]

Execute SQL queries to explore database content and gather evidence.

Use this when you need to:

- Discover possible values in a column (e.g., DISTINCT values)

- Verify data formats or patterns

- Check relationships between tables

- Gather sample data to understand the database
\\

**Exploration Guidelines**:

- Use LIMIT to restrict output when exploring specific values or samples.

- If you need to understand data distribution (e.g., range, distinct values), you may omit LIMIT. For large results (>30 rows), we will report: max value, min value, data format, and distinct values.
\\
            
**Format**: Start with [EXPLORE] tag, then write SQL queries with comments:

\verb|```|

[EXPLORE]

-- Purpose: Check available product categories

SELECT DISTINCT category FROM products LIMIT 10;

-- Purpose: Verify date format

SELECT date\_column FROM orders LIMIT 5;

\verb|```|
            
**Important**: After exploration, please use [REFINE] to analyze the results before generating SQL.
\\
            
\#\# [REFINE]

Analyze exploration results, update your understanding, and plan the next steps.

Use this to:

- Summarize what you learned from exploration and the remaining problems

- Update your logical plan

- Plan the SQL query structure (JOINs, filters, aggregations, etc.)

- Decide if more exploration is needed or if you're ready to generate SQL
\\

**Format**: Start with [REFINE] tag, then provide structured reasoning:

\verb|```|

[REFINE]

\#\#\# Findings from Exploration:

- [Summarize key discoveries]

\#\#\# Updated Understanding:

- [How this changes your approach]

\#\#\# Query Plan:

- [Step-by-step plan for the SQL query]

\#\#\# Next Action:

- [EXPLORE more] OR [Generate SQL]

\verb|```|
\\

\#\# [SQL]

Generate the final SQL query.

Use this when you are confident about the query logic.
\\

**Format**: Start with [SQL] tag, then provide the query:

[SQL] \verb|```|sql <Your SQL query> \verb|```|
\\

\#\# [CONFIRM]

Confirm the logic of the generated SQLs and the final result after SQL execution.

Use this ONLY after [SQL] execution returns a satisfactory result.

**Format**: [CONFIRM] <Brief description of what the query does>
\end{promptbox}

\begin{promptbox}[Evidence Linking]
You are an expert Data Analyst Assistant supporting a Text-to-SQL system.

We have a User Query and an External Knowledge Document (Markdown format) that contains business rules, calculation logic, or data dictionary definitions.

Your task is to **extract** every piece of information from the document that is relevant to the User Query.
\\

\#\#\# Input Information

\quad- **User Query**: \{query\}

\quad- **Knowledge File Name**: \{knowledge\_file\_name\}

\quad- **Original Knowledge Content**:

\verb|```|markdown

\{knowledge\_content\}

\verb|```|
\\

\#\#\# Extraction Instructions (CRITICAL)

1. **Goal: High Recall (Better Safe Than Sorry).**

\quad- If any section, paragraph, definition, description, entity code, formula, or table row is **potentially** related to the entities, metrics, conditions, constraints, or logic in the query (even slightly), **KEEP IT**.

\quad- Do NOT try to be concise. We prefer extra context over missing information.

\quad- Only remove content that is obviously and strictly irrelevant (e.g., legacy codes not mentioned, definitions of completely unrelated departments).

2. **Maintain Context \& Integrity.**

\quad- Do NOT pick out single words or fragmented sentences.

\quad- Keep entire paragraphs, list items, or table rows to ensure the context remains readable and authentic.

\quad- If a calculation rule depends on previous lines (like a variable definition), include those lines too.

3. **Do Not Rewrite.**

\quad- Do NOT summarize, paraphrase, or change the original text. **Copy and paste** the relevant sections exactly as they appear in the source.
\\

\#\#\# Output

Output ONLY the extracted markdown content below, without any introductory or concluding text.
\end{promptbox}

\begin{promptbox}[Answer Selection]
You are a Senior Data Architect acting as a Judge. You are provided with a User Question, the Database Schema, and several Candidate Solutions generated by an AI agent.

Each candidate consists of:

1. **The Execution Strategy**: The logic derived after exploring the database (identifying specific tables, columns, and values).

2. **The Final SQL**: The query implementation (We have varified that the SQL is executable).

**YOUR GOAL**: Identify the SINGLE best candidate that is most likely to execute correctly and return the accurate answer.
\\

*** DATABASE SCHEMA ***

\{schema\}
\\

*** USER QUESTION ***

\{question\}
\\

*** CANDIDATES ***

\{candidates\}
\\

*** EVALUATION CRITERIA (Prioritize in this order) ***
            
1. **Specificity of Evidence (The "Verified" Test)**:

\quad- **Favor** candidates where the Strategy explicitly lists *verified values* found during exploration.

\quad- **Reject** candidates with vague strategies (e.g., "Filter by population metric" without stating *which* metric ID).       
2. **Entity Isolation (The "Explosion" Test)**:

\quad- Look at the Schema. If a table contains mixed data types (e.g., `MetricID`, `EventType`, `Year`), the SQL **MUST** filter for a specific value.

\quad- **Reject** candidates that aggregate a Fact/Event table without a `WHERE` clause filtering for the specific metric/type (this leads to wrong sums).
            
3. **Logic Robustness (The "Safety" Test)**:

\quad- **Ratios**: The SQL should handle zero denominators (e.g., `WHERE denom > 0` or `NULLIF`).

\quad- **Joins**: If the task involves multiple independent event tables (e.g., Sends, Opens), **Favor** candidates using `UNION ALL` or `FULL JOIN` strategies over simple `INNER/LEFT JOIN` which might lose data.
            
4. **Consistency**:

\quad- The SQL must strictly follow the Strategy. If the Strategy says "Filter X" but SQL does not, reject it.
\\

*** OUTPUT INSTRUCTION ***

1. Analyze each candidate one by one based on the criteria above.

2. Compare the candidates (both the strategy and the SQL) to point out if some of them miss necessary filters (Entity Isolation) or lacks specific verified details.

3. Select the best candidate.

4. Output the chosen file name in this format:

\verb|```|plaintext

xxx.sql

\verb|```|
\end{promptbox}

\section{Tip Library Details}
\label{appx:library}
This document presents the complete tip library $\mathcal{M}$ used for SQL generation guidance (Section \ref{sec:derterminitic_retrieval}). The library contains 49 tips organized into 14 categories. We use DeepSeek-V3.2 to analyze a randomly selected 10\% of BIRD-train cases to extract these tips. Each tip is formatted as it appears in the actual case prompts: [TIP\_ID] Title followed by Description.

\subsection{Categories Overview}

\begin{itemize}[leftmargin=*,itemsep=0pt,parsep=0.1em,topsep=0.1em,partopsep=0.1em]
\item evidence\_enforcement: Tips for correctly implementing evidence
\item string\_matching: Tips for choosing = vs LIKE
\item output\_columns: Tips for selecting correct output columns
\item table\_selection: Tips for choosing tables and joins
\item column\_interpretation: Tips for understanding column names
\item schema\_grounding: Tips for mapping logical concepts to schema
\item join\_strategy: Tips for implementing joins
\item filter\_implementation: Tips for implementing WHERE conditions
\item aggregation: Tips for aggregation and calculations
\item sorting\_limiting: Tips for ORDER BY and LIMIT
\item multi\_step\_logic: Tips for subqueries and CTEs
\item sql\_syntax: Tips for SQL syntax correctness
\item common\_pitfalls: Tips for avoiding common errors
\item advanced\_patterns: Tips for advanced SQL patterns
\end{itemize}

\subsection{Complete Tip Listing}

\subsubsection{Category: Evidence Enforcement}
\mbox{} \\ \indent
\textbf{[TIP001]} Evidence-Specified Column Must Be Used Exactly

When evidence specifies a column name, explore using that exact column. Do NOT substitute with similar columns.

\textbf{[TIP002]} Evidence-Specified Formula Must Be Implemented Exactly

When evidence provides a calculation formula, explore implementing it EXACTLY as specified. Do NOT use built-in functions that seem equivalent.

\textbf{[TIP003]} Evidence-Specified Value Must Be Used Exactly

When evidence provides value mappings, explore using the EXACT database value, not the English translation or interpretation.

\textbf{[TIP004]} Evidence Clarifies Ambiguous Terms

When evidence clarifies which column or table an ambiguous term refers to, explore following the evidence interpretation.

\textbf{[TIP044]} Evidence Formula Preservation - No Rephrasing Allowed

When evidence provides a formula or calculation method, explore preserving it EXACTLY. Do NOT rephrase, simplify, or use 'equivalent' built-in functions.

\textbf{[TIP045]} WHERE Clause from Evidence - Extract Exact Conditions

When evidence specifies how to filter data (column names, values, or conditions), explore extracting and implementing these EXACTLY.

\subsubsection{Category: String Matching}
\mbox{} \\ \indent
\textbf{[TIP005]} Use Exact Match (=) When Value Is Known

When question or evidence provides a specific value, explore using exact match (=). Only use LIKE for partial matching.

\textbf{[TIP006]} Use LIKE Only for Partial Matching

Explore using LIKE only when question explicitly asks for partial matching (contains, starts with, ends with).

\textbf{[TIP007]} After Exploration, Use Exact Values Not Patterns

After exploring to find possible values, use exact match (=) with the found value to explore, not LIKE pattern.

\textbf{[TIP025]} Choose String Matching Strategy Based on Question Intent

Explore choosing between exact match (=), prefix match (LIKE 'X\%'), suffix match (LIKE '\%X'), or contains (LIKE '\%X\%') based on question.

\subsubsection{Category: Output Columns}
\mbox{} \\ \indent
\textbf{[TIP008]} Exclude ORDER BY Columns from SELECT Unless Requested

Explore excluding columns used only for sorting (ORDER BY) from SELECT unless explicitly requested in question.

\textbf{[TIP009]} Return Columns in Exact Order Mentioned in Question

Explore returning output columns in the EXACT order they appear in the question. First mentioned → first in SELECT.

\textbf{[TIP010]} Exclude Filter Columns from SELECT Unless Requested

Explore excluding columns used only for filtering (WHERE) from SELECT unless explicitly requested.

\textbf{[TIP011]} For 'How Many' Questions, Return COUNT Only

When question asks 'how many', explore returning only COUNT, not individual records or other columns.

\textbf{[TIP043]} SELECT Clause Precision - Return ONLY Requested Columns

Explore ensuring the SELECT clause returns ONLY the columns explicitly requested in the question, in the exact order mentioned.

\subsubsection{Category: Table Selection}
\mbox{} \\ \indent
\textbf{[TIP012]} Check If Single Table Contains All Required Information

Before exploring JOIN, verify if all required information exists in a single table. Prefer exploring single-table solutions.

\textbf{[TIP013]} JOIN Only When Information Spans Multiple Tables

Explore using JOIN only when question explicitly needs information from multiple entities that exist in different tables.

\textbf{[TIP014]} Avoid Over-Engineering with Unnecessary Complexity

Explore avoiding JOINs, subqueries, or conditions that aren't needed. Use the simplest solution that answers the question.

\subsubsection{Category: Column Interpretation}
\mbox{} \\ \indent
\textbf{[TIP015]} Parenthetical Text in Column Names Is NOT a Filter Condition

When exploring, note that text in parentheses within column names (e.g., '(K-12)', '(Ages 5-17)') is part of the column name, NOT a filter condition.

\textbf{[TIP016]} Distinguish Column Name Context from Filter Conditions

When exploring, note that parenthetical text in column names provides context about the data, not additional filter conditions to apply.

\subsubsection{Category: Schema Grounding}
\mbox{} \\ \indent
\textbf{[TIP017]} Use Direct Name Matching for Column Identification

When exploring, if logical plan mentions a concept, look for columns with similar names first (case-insensitive, with variations).

\textbf{[TIP018]} Use Semantic Matching When Direct Names Don't Match

When exploring, if direct name matching fails, think about what the concept means and look for semantically related columns.

\textbf{[TIP019]} Always Verify Column Existence Before Using

When exploring, never assume a column exists. Always verify through exploration before using.

\subsubsection{Category: Join Strategy}
\mbox{} \\ \indent
\textbf{[TIP020]} Identify Join Keys from Schema Relationships

When exploring joins, look for foreign key relationships in schema. Common patterns: ID columns, Code columns, Name columns.

\textbf{[TIP021]} Verify Join Keys Have Matching Values

When exploring joins, verify that join keys have matching values in both tables and check for NULLs.

\textbf{[TIP022]} Choose Appropriate Join Type

When exploring joins, select join type based on whether you need all records from one/both tables or only matching records.

\textbf{[TIP042]} Prefix Columns with Table Alias When Joining

When exploring joins, prefix all columns with table alias to avoid ambiguous column errors.

\subsubsection{Category: Filter Implementation}
\mbox{} \\ \indent
\textbf{[TIP023]} Use Exact Value Filters When Value Is Specified

When question explicitly mentions a value, explore using exact match (=) for filtering.

\textbf{[TIP024]} Use Correct Numeric Comparison Operators

When exploring filters, translate question's numeric comparisons to correct operators. Pay attention to inclusive vs exclusive. BETWEEN is inclusive on both ends.

\subsubsection{Category: Aggregation}
\mbox{} \\ \indent
\textbf{[TIP026]} Choose Correct Aggregation Function

When exploring aggregations, select appropriate function based on what question asks: COUNT, SUM, AVG, MAX, MIN.

\textbf{[TIP027]} Protect Division Operations Against Zero and Use CAST

When exploring calculations with ratios or percentages, protect against division by zero and ensure float division with CAST.

\textbf{[TIP028]} Use GROUP BY When Aggregating Per Category

When exploring aggregations and question asks 'per category', 'by group', 'for each', use GROUP BY with the category column.

\textbf{[TIP029]} Use HAVING to Filter After Aggregation

When exploring aggregations, use HAVING clause to filter groups after aggregation. Use WHERE to filter rows before aggregation.

\textbf{[TIP046]} Precision Preservation - No Arbitrary Rounding

When exploring calculations, do NOT apply ROUND() unless the question explicitly specifies precision requirements.

\textbf{[TIP047]} Ratio Calculation - Count All Occurrences

When exploring ratio or percentage calculations, count ALL occurrences (including duplicates) unless the question explicitly asks for DISTINCT counts.

\textbf{[TIP048]} Percentage Calculation - CAST to REAL and Multiply by 100

When exploring percentage calculations, always CAST the numerator to REAL before division, then multiply by 100.

\textbf{[TIP049]} DISTINCT Usage - List vs Count

When exploring, use DISTINCT when listing entities (names, IDs, items) to avoid duplicates. Do NOT use DISTINCT when counting occurrences for statistical purposes unless explicitly asked for unique counts.

\subsubsection{Category: Sorting and Limiting}
\mbox{} \\ \indent
\textbf{[TIP030]} Implement ORDER BY Based on Question Keywords

When exploring sorting, translate question's keywords to correct ORDER BY direction: ASC for lowest/first, DESC for highest/last.

\textbf{[TIP031]} Use LIMIT for Top N or Single Extreme Value

When exploring and question asks for 'top N', 'first N', or single extreme value, use LIMIT with ORDER BY.

\textbf{[TIP032]} Filter Out NULLs When Finding Extreme Values

When exploring extreme values with ORDER BY, add WHERE column IS NOT NULL to avoid NULL results.

\subsubsection{Category: Multi-Step Logic}
\mbox{} \\ \indent
\textbf{[TIP033]} Use Subquery for Dependent Steps

When exploring and logical plan has dependent steps (step 2 needs result from step 1), explore using subquery or CTE.

\textbf{[TIP034]} Use CTE for Complex Multi-Step Logic

When exploring complex queries with multiple dependent steps, explore using CTE (WITH clause) for clarity.

\subsubsection{Category: SQL Syntax}
\mbox{} \\ \indent
\textbf{[TIP035]} Use Double Quotes for Identifiers with Spaces or Special Characters

When exploring, note that column/table names with spaces, special characters, or reserved words must be quoted with double quotes.

\textbf{[TIP036]} Use Aliases for Calculated Columns

When exploring calculations, give meaningful aliases to calculated columns, aggregations, and complex expressions using AS.

\textbf{[TIP037]} Cannot Use Column Alias in WHERE Clause of Same Level

When exploring, note that column aliases defined in SELECT cannot be used in WHERE clause of the same query level. Use subquery or repeat expression.

\subsubsection{Category: Common Pitfalls}
\mbox{} \\ \indent
\textbf{[TIP038]} Never Hard-Code Values Discovered in Exploration

When exploring, do NOT hard-code values you discovered unless they are explicitly mentioned in the question.

\textbf{[TIP039]} Handle NULL Values Appropriately

When exploring, be aware of NULL values in aggregations, comparisons, and joins. Filter or handle them explicitly.

\textbf{[TIP040]} Avoid Integer Division - Use CAST for Float Results

When exploring calculations in SQLite, note that dividing two integers gives integer result. Use CAST(x AS REAL) to get float result.

\textbf{[TIP041]} Apply Entity Isolation for Mixed-Data Tables

When exploring tables with mixed entity types (e.g., different metrics in same table), filter by entity type before aggregating.

\subsubsection{Category: Advanced Patterns}
\mbox{} \\ \indent
\textbf{[TIP043]} Use Conditional Aggregation with CASE

When exploring aggregations of different subsets in one query, use CASE inside aggregate functions.

\textbf{[TIP044]} Use Window Functions for Ranking

When exploring ranking, row numbering, or running calculations, use window functions (RANK, ROW\_NUMBER, etc.).

\textbf{[TIP045]} Find Top N Per Group with Window Functions

When exploring top N records per category, use ROW\_NUMBER() with PARTITION BY, then filter.

\subsection{Tip Retrieval Relationships}

The tip retrieval system uses rule-based keyword matching to select relevant tips for each question.

\subsubsection{Universal Tips (Always Selected)}
\mbox{} \\ \indent
The following tips are included in every query regardless of content:
\textbf{TIP009}, \textbf{TIP019}, \textbf{TIP035}, and \textbf{TIP038}.

\subsubsection{Evidence-Based Retrieval}

Tips selected based on patterns detected in the evidence field:

\begin{itemize}[leftmargin=*,itemsep=0pt,parsep=0.1em,topsep=0.1em,partopsep=0.1em]
\item Evidence contains phrases like "can be represented as [column]=" or "[column]=[value]" (excluding "refers to") \textrightarrow\ \textbf{TIP001}
\item Evidence contains formulas with arithmetic operators, or phrases like "sum of count" or "average =" \textrightarrow\ \textbf{TIP002, TIP027, TIP040}
\item Evidence contains "stands for" or "means" (with quoted values), or patterns like "'value' stands for/means/is/are" \textrightarrow\ \textbf{TIP003}
\item Evidence contains "refers to" \textrightarrow\ \textbf{TIP004}
\end{itemize}
\subsubsection{Question-Based Retrieval}

Tips selected based on keywords and patterns in the question text:

\begin{itemize}[leftmargin=*,itemsep=0pt,parsep=0.1em,topsep=0.1em,partopsep=0.1em]
\item Question contains "with", "where", or "whose" followed by a quoted value \textrightarrow\ \textbf{TIP005, TIP023}
\item Question contains: highest, lowest, top, bottom, maximum, minimum, best, worst, most, least, order by, sort \textrightarrow\ \textbf{TIP008, TIP030, TIP031, TIP032}
\item Question mentions multiple outputs connected by "and" (e.g., "name and age"), but does NOT start with "how many" \textrightarrow\ \textbf{TIP009}
\item Question starts with "what", "list", "name", or "show" and contains "where", "with", "whose", or "that have" \textrightarrow\ \textbf{TIP010}
\item Question starts with "how many" (without "what" or "which"), or contains "list" + "lowest" + "amount" \textrightarrow\ \textbf{TIP011}
\item Question or evidence contains parentheses \textrightarrow\ \textbf{TIP015, TIP016}
\item Question contains: more than, less than, greater than, between, at least, at most, above, below, over, under, exceed \textrightarrow\ \textbf{TIP024}
\item Question contains: total, sum, average, avg, count, maximum, minimum, aggregate, group by \textrightarrow\ \textbf{TIP026, TIP028}
\item Question contains aggregation keywords (average, total, sum, count) combined with comparison keywords (more than, less than, greater, between) \textrightarrow\ \textbf{TIP029}
\item Question contains: calculate, ratio, percentage, average \textrightarrow\ \textbf{TIP036, TIP046}
\item Question contains: ratio, percentage, percent, proportion, rate \textrightarrow\ \textbf{TIP047, TIP048}
\item Question starts with: how many, what is, what are, which, list all, list the \textrightarrow\ \textbf{TIP049}
\item Question contains: contain, include, start with, end with, like \textrightarrow\ \textbf{TIP006}
\item Question contains: name, title, description \textrightarrow\ \textbf{TIP025}
\item Question contains: null, missing, empty \textrightarrow\ \textbf{TIP039}
\end{itemize}

\subsubsection{Logical Plan-Based Retrieval}

Tips selected based on patterns in the logical plan structure:

\begin{itemize}[leftmargin=*,itemsep=0pt,parsep=0.1em,topsep=0.1em,partopsep=0.1em]
\item Plan contains "join", or mentions 4+ tables, or question contains patterns like "X and their/its Y" \textrightarrow\ \textbf{TIP012, TIP013, TIP020, TIP021, TIP022, TIP042}
\item Plan contains both "subquery" and "nested", or has 4+ occurrences of "join" \textrightarrow\ \textbf{TIP014}
\item Plan has 4+ steps, or contains both "subquery" and "nested", or uses CTE pattern (WITH...AS...SELECT) \textrightarrow\ \textbf{TIP033, TIP034}
\item Plan contains "case when", or contains both "different" and "aggregate" \textrightarrow\ \textbf{TIP043}
\item Plan contains: rank, ranking, row number, top n per, per group, partition \textrightarrow\ \textbf{TIP044, TIP045}
\end{itemize}

\subsubsection{Schema-Based Retrieval}

Tips selected based on database schema characteristics:

\begin{itemize}[leftmargin=*,itemsep=0pt,parsep=0.1em,topsep=0.1em,partopsep=0.1em]
\item Schema contains entity type columns (metricid, metric\_id, event\_\\
type, entity\_type, data\_type) and question contains filtering keywords (specific, certain, particular, only, filter by type, where type) \textrightarrow\ \textbf{TIP041}
\end{itemize}

\section{Case Study}
\label{appx:case_study}
To demonstrate the effectiveness of agentic exploration in SQL generation, we present a representative case from the Spider 2.0-Snow benchmark. This case (\textbf
{sf\_bq028}) illustrates how agentic exploration enables the model to discover critical data quality issues and adapt its strategy accordingly, leading to successful query generation where the baseline method failed.

\subsection{Case sf\_bq028: Data Quality Discovery Through Exploration}

\subsubsection{Task Description}
Considering only the latest release versions of NPM packages, which packages are the top 8 most popular based on the Github star number, as well as their versions?

\subsubsection{Database Schema}

The task involves three tables from the \texttt{DEPS\_DEV\_V1} database:

\paragraph{Table: PROJECTS}
\begin{itemize}[leftmargin=2em]
    \item \texttt{Name} (TEXT): Project identifier (e.g., 'steven-tey/dub')
    \item \texttt{Type} (TEXT): Project type (e.g., 'GITHUB', 'GITLAB')
    \item \texttt{StarsCount} (NUMBER): Number of stars for the project
    \item \texttt{SnapshotAt} (NUMBER): Timestamp of data snapshot
\end{itemize}

\paragraph{Table: PACKAGEVERSIONS}
\begin{itemize}[leftmargin=2em]
    \item \texttt{Name} (TEXT): Package name (e.g., '@dub/ui')
    \item \texttt{Version} (TEXT): Version string (e.g., '1.1.6')
    \item \texttt{System} (TEXT): Package system (e.g., 'NPM', 'PYPI')
    \item \texttt{VersionInfo} (VARIANT): JSON containing \texttt{IsRelease} (boolean) and \texttt{Ordinal} (number)
    \item \texttt{UpstreamPublishedAt} (NUMBER): Publication timestamp (often NULL)
    \item \texttt{SnapshotAt} (NUMBER): Timestamp of data snapshot
\end{itemize}

\paragraph{Table: PACKAGEVERSIONTOPROJECT}
\begin{itemize}[leftmargin=2em]
    \item \texttt{System} (TEXT): Package system
    \item \texttt{Name} (TEXT): Package name
    \item \texttt{Version} (TEXT): Package version
    \item \texttt{ProjectType} (TEXT): Type of linked project (e.g., 'GITHUB')
    \item \texttt{ProjectName} (TEXT): Name of linked project
    \item \texttt{RelationType} (TEXT): Relationship type (e.g., 'SOURCE\_REPO \\
    \_TYPE')
\end{itemize}

\subsubsection{Direct SQL Generation Result \& Analysis}

The baseline method directly assumed that \texttt{UpstreamPublishedAt} could be used to identify the latest release version of each NPM package.

\paragraph{Generated SQL strategy:}
\begin{itemize}[leftmargin=2em]
    \item Filter \texttt{PACKAGEVERSIONS} for \texttt{System = 'NPM'} and \texttt{VersionInf \\
    o:IsRelease = TRUE}
    \item Use \texttt{ROW\_NUMBER() OVER (PARTITION BY "Name" ORDER BY "UpstreamPublishedAt" DESC)} to find latest release
    \item Filter \texttt{WHERE "UpstreamPublishedAt" IS NOT NULL}
\end{itemize}

Critical error: The baseline SQL included the condition:
\begin{lstlisting}[breaklines=true,basicstyle=\small\ttfamily]
WHERE "UpstreamPublishedAt" IS NOT NULL
\end{lstlisting}

Why this failed: This assumption was made without verification. The baseline had no way to know whether \texttt{UpstreamPublishedAt} was actually populated for NPM release versions. As it turns out, this field is NULL for the majority of NPM releases, causing the query to miss most packages and return incorrect results.

\paragraph{Result} \emph{Failed (Score: 0)}

\subsubsection{\TheName{} Result \& Analysis} \mbox{} \\ \indent
\textbf{Exploration Round 1: Timestamp Availability Check}

Query 1: Check if UpstreamPublishedAt is populated for release versions
\begin{lstlisting}[breaklines=true,basicstyle=\small\ttfamily]
SELECT "UpstreamPublishedAt", "VersionInfo" 
FROM "PACKAGEVERSIONS" 
WHERE "System" = 'NPM' AND "VersionInfo":IsRelease = true 
LIMIT 10;
\end{lstlisting}

Result: All 10 samples showed \texttt{UpstreamPublishedAt = None}

% \vspace{0.5em}
Query 2: Check if VersionInfo contains alternative timestamp
\begin{lstlisting}[breaklines=true,basicstyle=\small\ttfamily]
SELECT "VersionInfo" 
FROM "PACKAGEVERSIONS" 
WHERE "System" = 'NPM' AND "VersionInfo":IsRelease = true 
LIMIT 10;
\end{lstlisting}

Result: VersionInfo contains \texttt{IsRelease} boolean and \texttt{Ordinal} number, but no timestamp

% \vspace{0.5em}
Query 3: Check SnapshotAt as fallback
\begin{lstlisting}[breaklines=true,basicstyle=\small\ttfamily]
SELECT "SnapshotAt", "VersionInfo", "UpstreamPublishedAt" 
FROM "PACKAGEVERSIONS" 
WHERE "System" = 'NPM' AND "VersionInfo":IsRelease = true 
AND "UpstreamPublishedAt" IS NULL 
LIMIT 10;
\end{lstlisting}

Result: SnapshotAt is consistent across records (system snapshot time, not release time)

% \vspace{0.5em}
\emph{Key Discovery:} \texttt{UpstreamPublishedAt} is NULL for NPM releases, so it cannot be used!

% \vspace{0.5em}
Updated strategy:
\begin{quoting}[leftmargin=2em]
Evidence: \texttt{UpstreamPublishedAt} is NULL for all NPM release versions, so cannot be used for timestamp. \texttt{VersionInfo} contains \texttt{IsRelease} boolean and \texttt{Ordinal} number but no timestamp. \\
Updated Plan: Use \texttt{VersionInfo:Ordinal} to find latest release per package via \texttt{ROW\_NUMBER() OVER (PARTITION BY "Name" ORDER BY "VersionInfo":Ordinal DESC)}.
\end{quoting}

% \vspace{0.5mm}
\textbf{Exploration Round 2: Quantify the Impact}

Query: Count NULL UpstreamPublishedAt in releases
\begin{lstlisting}[breaklines=true,basicstyle=\small\ttfamily]
SELECT COUNT(*) AS null_count 
FROM "PACKAGEVERSIONS" 
WHERE "System" = 'NPM' AND "VersionInfo":IsRelease = true 
AND "UpstreamPublishedAt" IS NULL;
\end{lstlisting}

Result: \texttt{null\_count = 3,935,243} (out of $\sim$9M total NPM releases)

% \vspace{0.5em}
\textbf{Critical Finding:} 43.7\% of NPM releases have NULL UpstreamPublishedAt! (The baseline's assumption would have excluded nearly half of all packages.)

% \vspace{0.5em}
\textbf{Exploration Round 3: Verify Alternative Approach}

Query 1: Verify VersionInfo structure
\begin{lstlisting}[breaklines=true,basicstyle=\small\ttfamily]
SELECT "VersionInfo" 
FROM "PACKAGEVERSIONS" 
WHERE "System" = 'NPM' AND "VersionInfo":IsRelease = TRUE 
LIMIT 5;
\end{lstlisting}

Result: Confirmed \texttt{Ordinal} field exists and can be used for ordering

% \vspace{0.5em}
Query 2: Verify project name format for joins
\begin{lstlisting}[breaklines=true,basicstyle=\small\ttfamily]
SELECT DISTINCT "ProjectName" 
FROM "PACKAGEVERSIONTOPROJECT" 
WHERE "ProjectType" = 'GITHUB' 
LIMIT 5;
\end{lstlisting}

Result: Confirmed format matches \texttt{PROJECTS.Name} (e.g., 'steven-tey/dub')

% \vspace{0.5em}
\textbf{Final SQL Generation.}
Based on verified evidence, the model generated:
\begin{lstlisting}[breaklines=true,basicstyle=\small\ttfamily]
WITH LatestReleases AS (
    SELECT 
        "Name",
        "Version",
        ROW_NUMBER() OVER (
            PARTITION BY "Name" 
            ORDER BY "VersionInfo":Ordinal DESC
        ) AS version_rank
    FROM "PACKAGEVERSIONS"
    WHERE 
        "System" = 'NPM' 
        AND "VersionInfo":IsRelease = true
),
PackageProjects AS (
    SELECT DISTINCT
        lr."Name",
        lr."Version",
        pvp."ProjectName"
    FROM LatestReleases lr
    JOIN "PACKAGEVERSIONTOPROJECT" pvp
        ON lr."Name" = pvp."Name" 
        AND lr."Version" = pvp."Version"
        AND pvp."ProjectType" = 'GITHUB'
    WHERE lr.version_rank = 1
)
SELECT 
    pp."Name" AS PackageName,
    pp."Version",
    p."StarsCount"
FROM PackageProjects pp
JOIN "PROJECTS" p
    ON pp."ProjectName" = p."Name"
    AND p."Type" = 'GITHUB'
ORDER BY p."StarsCount" DESC
LIMIT 8;
\end{lstlisting}

\paragraph{Result} \emph{Success (Score: 1)}

\subsubsection{Key Insights}

Why exploration was essential:
\begin{enumerate}[leftmargin=1.5em]
    \item Data Quality Discovery: Exploration revealed that \texttt{UpstreamPub \\
    lishedAt} is NULL for 43.7\% of NPM releases---a critical data quality issue that the baseline could not detect.
    \item Alternative Strategy: After discovering the NULL issue, exploration identified \texttt{VersionInfo:Ordinal} as a reliable alternative for version ordering.
    \item Verification Before Commitment: The model verified the alternative approach before generating the final SQL, ensuring correctness.
\end{enumerate}

% \vspace{0.5em}
This case demonstrates that exploration is not guessing---it's systematic data quality verification that enables the model to adapt its strategy based on actual database characteristics.

\end{document}